# Dose Delivery Concept and Instrumentation

S. Giordanengo[1] and M. Donetti[2]
[1]Istituto Nazionale di Fisica Nucleare (INFN), Torino, Italy
[2]Centro Nazionale di Adroterapia Oncologica (CNAO), Pavia, Italy

**Abstract**

Radiation therapy aims to deliver the prescribed amount of dose to a tumour at the same time as sparing the surrounding tissues as much as possible. In charged particle therapy, delivering the prescribed dose is equivalent to delivering the prescribed number of ions of a given energy at each position of the irradiation field. The accurate delivery is committed to a dose delivery (DD) system that shapes, guides and controls the beam before the patient entrance. Most of the early DD systems provided uniform lateral dose profiles by using different devices, mainly patient-specific, placed in the beam line to shape the three-dimensional final target dose. More recently, systems that provide highly conformal dose distributions using thousands of narrow beams at well-defined energy were developed which feature advanced scanning magnets and real-time beam monitors, without patient-specific hardware. This lecture will cover the general dose delivery concept as well as the different DD instrumentations depending mainly on the beam delivery technique and on the particle and accelerator types. Some characteristic worldwide DD and beam monitor systems will be mentioned.

## 1    Introduction

In charged particle therapy the presence of a dose delivery (DD) system is mandatory to deliver the dose as prescribed. In general this translates into controlling the beam characteristics, such as the number of particles, in the expected position and with a defined spatial distribution. The DD system connects the accelerator to the patient (see Fig. 1(b)) and operates on the charged particle beams to provide the 'patient-specific' beam, based on clinical requirements. In fact, at the highest level, the overall goal of radiotherapy is to deliver the specific amount of dose to the target and in an acceptable time.

The delivery technique together with the beam characteristics provided by the accelerator are the essential information and constraints used by the treatment planning system (TPS) to compute the optimum treatment (i.e. the list of parameters to set up the accelerator and the dose delivery instrumentation) in order to meet the prescriptions defined by radiation oncologists and medical physicists. The TPS uses computer algorithms and radiobiological models to simulate the interactions between the charged particle beams and the patient's anatomy and to determine the spatial distribution of the radiation dose provided by known beam characteristics.

The charged particle beam, accelerated by either a cyclotron or a synchrotron, when it reaches the treatment room, is quasi-monoenergetic and the transverse profile can be approximated to a Gaussian with a width of the order of a centimetre or less. With such a beam, the dose delivered to a tissue would be initially fairly constant, with a sharp peak toward the end of travel of the ions (Bragg peak). Since the depth of the Bragg peak increases with the particle energy, the beam has to be spread transversely and modulated in energy to become clinically useful. Thus, specific dose delivery instrumentation is mandatory to perform this task.

For deep-seated tumours the treatment is usually composed of two or more irradiation fields to minimize the dose in surrounding organs, especially in the tissues passed through by the beam to reach the tumour, i.e. in the entrance volume. Radiation oncologists and medical physicists work together to define the desired treatment outcome, i.e. doses and constraints, which is then converted into beam parameters required by the therapy machine to deliver the dose. From the DD system and accelerator point of view, each irradiation field is a new and independent delivery, so, in the following, we will refer to treatment as a single-field irradiation.

As sketched in Fig. 1(a), the pristine particle beams, provided by any accelerator, are too small and narrow compared with the tumour dimension. The beam has to be modelled or deviated in order to conform the dose to the target volume (Fig. 1(a)). Some DD components could be in the accelerator vacuum pipe while others are in air, hosted in the very last part of the beam transport line called the nozzle.

The DD instrumentation depends on the nominal beam characteristics available at the DD system entrance (i.e. accelerator type and performance) and on the required beam specifications, which depend on the dose delivery technique and on the specific patient and target. Two general dose delivery techniques have been developed and used to deliver charged particle radiation therapy: the passive scattering and the scanning (or dynamic) beam widening [1–3]. We will see that specific DD systems exist for each combination of accelerator, particle and beam delivery technique.

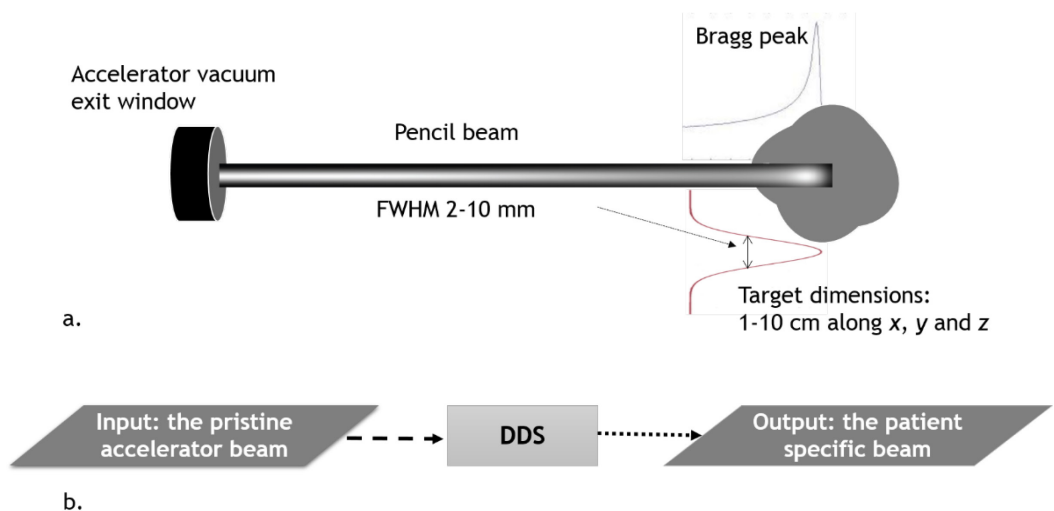

a.

b.

**Fig. 1:** (a) Sketch of a charged particle beam from the accelerator vacuum exit window to the target; (b) scheme of principles for the role of the dose delivery system (DDS).

One fundamental role of any DD system is the continuous check of beam parameters by means of dedicated beam monitors in order to interrupt the irradiation in case of values out of ranges. Real-time feedbacks to DD instrumentation can come from targets or from independent detectors as well as from the accelerator control system in order to guarantee the treatment accuracy and safety.

## 1.1 From the clinical requirements to treatment delivery

Patient treatment is a process that involves a large number of systems developed and interfaced with each other in order to achieve desired clinical requirements. Therefore, the latter are the starting specification that guides the construction of a particle therapy centre. The DD system allows measuring the clinical beam parameters, such as dose, dose rate, range, distal falloff, penumbra and degree of dose conformity, through the physical beam parameters such as beam current, beam energy, beam shape and size and beam position. A single clinical parameter can depend on many beam parameters and vice versa. For example, a change in the beam energy or range, position or shape could influence the dose uniformity or the unwanted dose outside the target.

The beam features are the result of a combination of the design of the accelerator, the beam transport and the beam-spreading mechanisms; tolerances associated with each beam parameter are important and must be derived from the maximum tolerable uncertainties of the overall treatment process.

The DD system is one among several other systems, listed in the following, mandatory to perform any kind of radiotherapy treatment:
- accelerator;
- beam transport lines;
- dose delivery system;
- patient positioning system;
- patient position verification system;
- treatment planning system;
- oncological information system;
- safety system.

By selecting the particle type, the accelerator technology and the dose delivery technique, different outcomes can be achieved in term of good coverage of the target volume, treatment time, treatment cost and patient throughput. Strengths and weaknesses exist for each system.

The basic clinical requirements in the design of a facility are the maximum tumour depth (typically about 30 cm) and dimension (more than 40 cm for cranio-spinal irradiation), the treatment duration (i.e. 2 Gy/min/l) and the desired dose uniformity in the target (typically better than 1–2%). As an example for carbon and proton beams, these requests lead to the parameters and field characteristics listed in Table 1.

**Table 1:** Main beam parameters and target characteristics in charged particle therapy

| Ion type | Energy range (MeV/u) | Flux ($N_{part}$/s) | Transversal beam position resolution (mm) | Transversal field size (cm min–max) | FWHM (cm $E_{min}$–$E_{max}$) |
|---|---|---|---|---|---|
| Proton | 70–250 | $10^9$–$10^{10}$ | ±1 | 1–40 | 2.3–0.7 |
| Carbon | 70–400 | $10^7$–$10^8$ | ±1 | 1–20 | 1–0.4 |

The bigger field size for proton beams is available with passive scattering techniques while for carbon ions $20 \times 20$ cm$^2$ is the recommended field dimension.

## 2    Dose delivery techniques

The delivery techniques that have been developed and used worldwide can be grouped under two broad categories: passive scattering and pencil beam scanning. Obviously each of these two categories has been implemented following different specific applications, but for sake of simplicity we will group them under the above classification: (1) 3D dose modulation with passive scattering techniques and (2) 3D dose modulation with pencil beam scanning techniques.

### 2.1    3D dose modulation with passive scattering techniques

The first charged particle delivery technique implemented has been the passive scattering. This method makes use of passive devices placed along the beam line that introduce scattering effects and energy degradation in order to spread out the Bragg peak transversally and along the beam direction; see Refs. [1–4].

This technique was taking advantage of the experience achieved with photon beams, known as conventional radiotherapy. In those applications, the photons (typically accelerated up to 18 MeV) are spread and collimated in the transverse plane by dose delivery devices placed in the accelerator nozzle. These devices are set to conform the irradiation field to the patient-specific case.

In the following sections we describe the use of the passive devices used in hadron therapy, such as: scatterers, range shifters, ridge filters, collimators, absorbers, apertures and range compensators.

### 2.1.1 Transversal beam modulation through beam spread and beam absorption

To increase the transverse dimensions of the pristine beam, one or two thin layers of high-$Z$ materials, like lead and tantalum, are placed in the beam line. Materials with large values of $Z$ are preferable because for a given energy loss the resulting multiple scattering angle is larger. Techniques using single or double scattering layers have been developed as shown in Fig. 2, reproduced from [5].

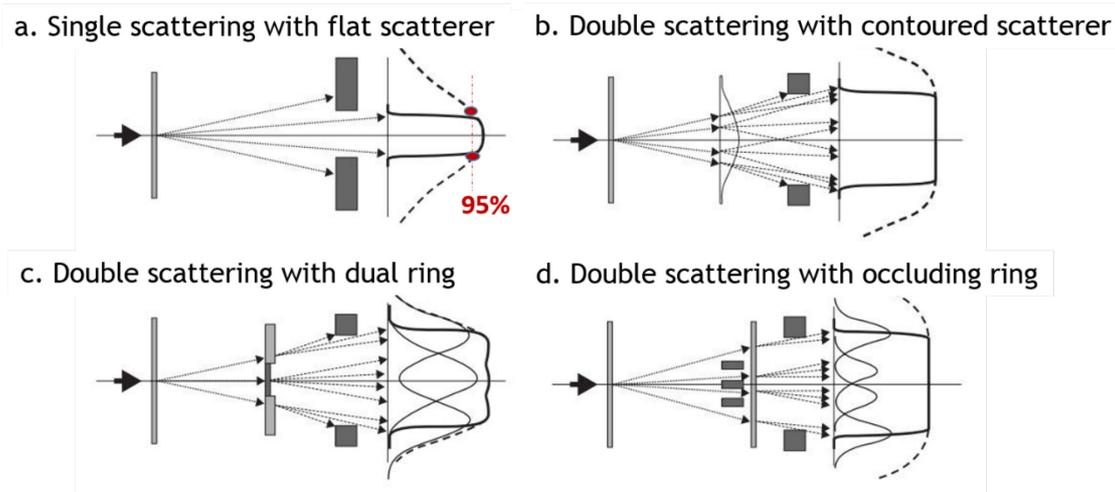

**Fig. 2:** (Reproduced from [5]) Schematic representation of the single-scattering technique using a flat scatterer (a) and double-scattering techniques using a contoured scatterer (b), dual ring (c) and occluding ring (d). Dashed lines, lateral profile without aperture; solid lines, with aperture.

To keep the dose uniformity within ±2.5%, a collimator (aperture) blocks the beam outside the 95% dose level so the useful part of the beam is only the 5%. Because of this low efficiency, requiring relatively large beam currents and generating a high production of secondary neutrons, a single flat scatterer is limited to small fields with a diameter typically not exceeding ~7 cm (mainly ocular tumours). To achieve a larger and uniform dose distribution, a second scatterer has to be placed in the beam line. This type of system is called a 'double-scattering system' and can be implemented with three different types of second scatterers as shown in Fig. 2: (b) the contoured, (c) the dual ring and (d) the occluding ring. The shape of a contoured scatterer, thick in the centre and thin on the outside, increases the efficiency because the central particles scatter more to the outside and the flat profile increases (Fig. 2(b)). Typically, a first flat scatterer spreads the beam onto the contoured scatterer (second scatterer) that flattens out the profile at some distance. The characteristics of the contoured scatterer are optimized in combination with the scattering power of the first scatterer to obtain the desired dose distribution. Efficiencies of up to 45% can be obtained, significantly larger than in single scattering.

Charged particles hitting the centre of the contoured scatterer lose more energy than those going through the thinner parts at the periphery. To avoid a concave distortion of the distal isodose plane, with the range increasing away from the beam axis, energy compensation is applied to the contoured scatterer. A high-$Z$ scattering material (lead, brass) is combined with a low-$Z$ compensation material (plastic). The thickness of the two materials is designed to provide constant energy loss, while maintaining the appropriate scattering power variation. The thickness of the high-$Z$ material decreases with distance from the axis, whereas the thickness of the compensating low-$Z$ material increases. Note that energy compensation will increase the total water equivalent thickness of the scatterer and the energy of the particles entering the nozzle needs to be increased to achieve the same range in the patient as with a single scatterer.

An alternative to the contoured scatterer is the dual-ring scatterer shown in Fig. 2(c) and also described in Ref. [5]. It consists of a central disc made of a high-$Z$ material (lead, tungsten) and a surrounding ring of a lower-$Z$ material (aluminium, Lucite). The physical thickness of the outer ring is chosen such that the energy loss is equal (or close) to the energy loss in the central disc. The first flat scatterer spreads the beam onto the dual-ring scatterer. The central disc produces a Gaussian-like profile, and the ring produces an annulus-shaped profile, and both combine to produce a uniform profile at the isocentre (Fig. 2(c)). Because of the interface between the two materials, the dose distribution is not perfectly flat.

The third type of second scatterer is the dual ring, which blocks instead of scattering the central protons outward (Fig. 2(d)). The 'hole' created in the fluence distribution is filled in by scattering through a flat, second scatterer. Larger field sizes can be obtained by not just blocking the centre but by adding one or more occluding rings. Optimization of the ring diameters and the first scatterer power results in a flat dose distribution. Because the protons are not redistributed but blocked, the efficiency of an occluding ring system is significantly lower than for the contoured scatterers. The energy loss is smaller though, because a relatively thin second scatterer foil is needed to spread out the beam.

All the double scattering systems are sensitive to beam alignment while the contoured and dual-ring scatterers also depend on the beam phase space (see Section 3).

Once the lateral beam spread is achieved, a field-specific aperture (Fig. 3(a)), called a collimator, is used to conform it laterally such that the dose matches the maximum beams-eye-view extent of the target volume. The 2D projection of the target along the beam direction is presented in Fig. 3(b). Alternatively, a multileaf collimator (MLC), made of tungsten leaves [5] can be used (Fig. 3(c)) as a dynamic aperture.

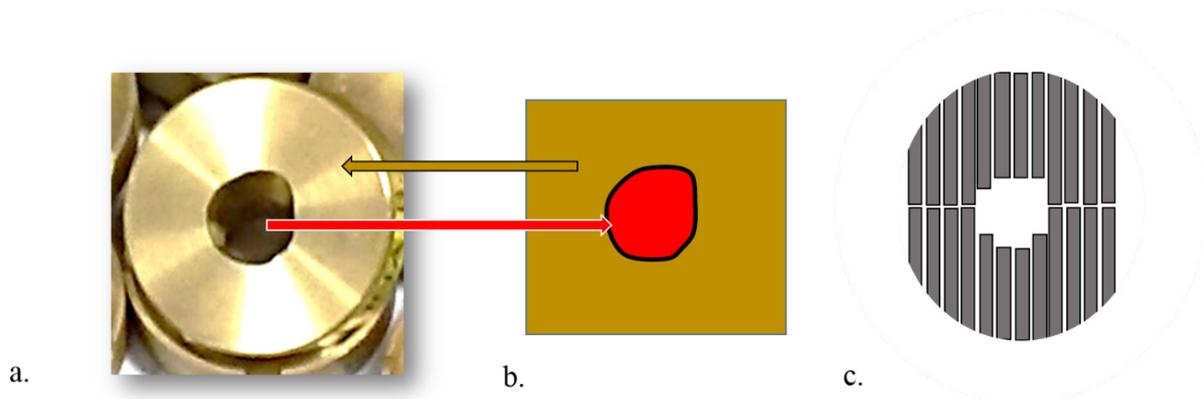

**Fig. 3:** (a) Example of brass field-specific aperture; (b) its 2D projection of the target; (c) scheme of a multileaf collimator.

Each delivered field requires apertures (or collimators) and range compensators to conform the dose to the target. The aperture absorbs the beam outside the target and conforms the beam laterally, while the range compensator is a patient-specific range shifter that conforms the dose delivered to the distal end of the target. It is also used to correct for patient surface irregularities and density heterogeneities in the beam path. Examples of lucite, wax and acrylic range compensators are shown in Fig. 4a, 4b and 5c respectively.

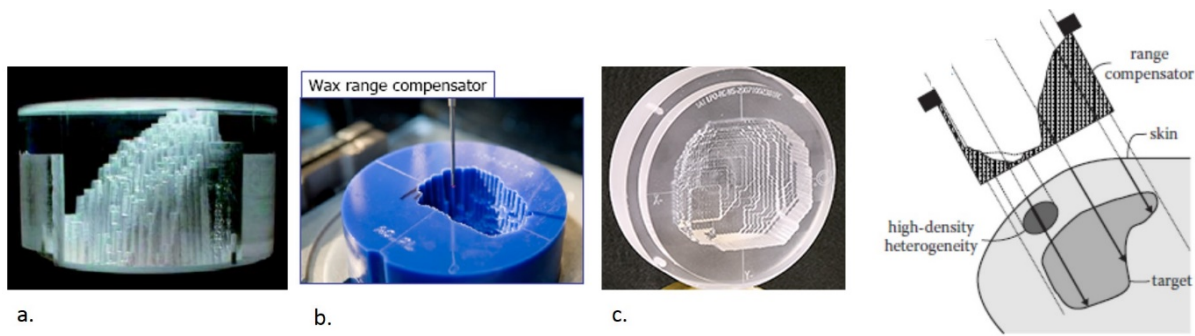

**Fig. 4**: Examples of range compensators made of lucite a); wax b) (photos by courtesy of MGH/LLUMC); acrylic c) (reproduced from http://www.oncolink.org/treatment/article.cfm?c=186&id=438) and a schematic representation of the application of a range compensator that compensates for the shape of the body entrance, the distal target shape, and inhomogeneities [5].

### 2.1.2    *Longitudinal beam spread by beam range modulation*

A uniform dose over the longitudinal extension of the tumour is obtained by modulating the beam range. The incident beam forms a flat dose region called the spread out Bragg peak (SOBP) by sequentially penetrating absorbers of variable thickness, e.g. via a range modulator. Each absorber contributes an individual pristine Bragg peak curve to the composite SOBP. A set of pristine peaks is delivered with decreasing depth and with reduced dose until the desired modulation is achieved. The result is the convolution of several Bragg peaks shifted in depth and with an appropriate weight, i.e. number of particles. Figure 5 shows a series of weighted pristine peaks as well as the resulting SOBP when these are superimposed.

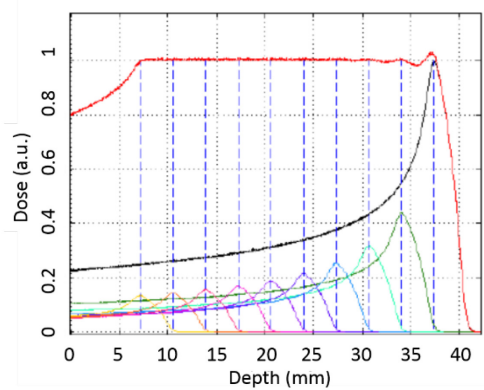

**Fig. 5:** Example of spread out Bragg peak (SOBP) to cover a 3 cm target extension in depth

The shape of the Bragg peak depends on the energy spread and scattering properties of the delivery system, so measured Bragg curves are used by TPS algorithms to determine the weights and the number of peaks necessary for each specific target volume treated in a specific facility.

The size of the SOBP is chosen to cover the largest extent in depth of the target volume. Since the SOBP size is constant over the entire target volume, in general, there is some pull back of the high-dose region into normal tissues proximal to the target volume.

### 2.1.2.1    *Energy degraders: range shifters, range modulator wheels and compensators*

Several different degraders have been developed and used either to change the fixed energy provided by the cyclotrons or to adapt the energy step and range provided by a given synchrotron. In Fig. 6, from the C. Ma and T. Lomax book [6], the following types are shown: (a) two or one adjustable wedges; (b) insertable slabs of graphite or Plexiglas, (c) rolled-up wedge; (d) insertable blocks with different thicknesses; (e) rotatable Plexiglas curved wedge; (f) adjustable multiwedge design.

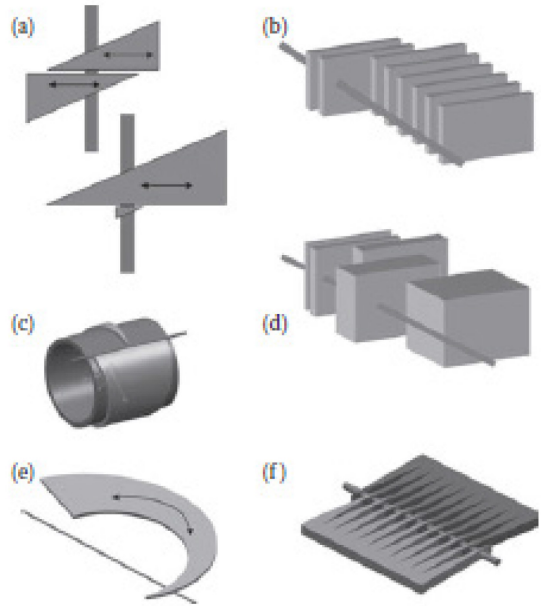

**Fig. 6:** Examples of energy degraders (reproduced from Ref. [6])

Different range modulator wheels, made of steps of varying thickness, are shown in Fig. 7. Each step corresponds to a pristine peak in the SOBP and the step thickness determines the range shift of that peak. When the wheel rotates in the beam, the steps are sequentially irradiated. The angular width of the step determines the number of protons hitting the step, and thus the weight of the pristine peak. By progressively increasing the step thickness while making the angular width smaller, a flat SOBP can be constructed. Like the range shifters used in energy stacking, modulator wheels are preferentially made of low-Z materials to limit scattering. Plastics (Plexiglas, Lexan) are often used, but for wheels that need to provide large range shifts and that are mounted in nozzles where space is limited, carbon and aluminium have been applied.

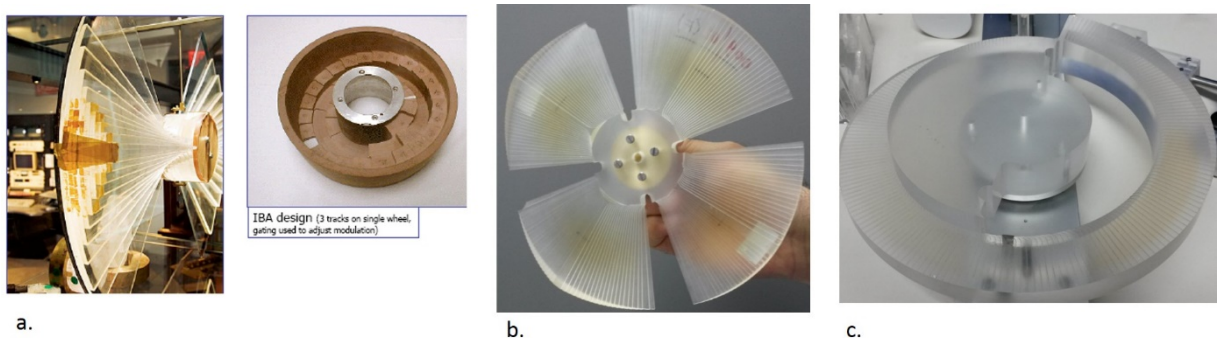

**Fig. 7:** Examples of range modulator wheels (a. and b. photos by courtesy of IBA)

This technique is not optimal in terms of the conformity of the dose deposition and the use of a gantry is recommended. Particles heavier than protons produce fragments and an undesired dose is delivered after the Bragg peaks. Moreover, it results in an additional dose to the patient due to the neutrons produced by the beam passing through the various components.

## 2.2    Pencil beam scanning techniques: 3D modulated scanning ion therapy

The pencil beam scanning technique exploits the physical proprieties of charged particles: a thin 'pencil' beam (typically 3–10 mm FWHM (full width at half maximum) at isocentre) coming directly from the beam line is transported to the patient to achieve small depositions of dose. The tumour is irradiated by the superposition of a sequence of beamlets, hereafter called spots, each delivering a defined number of particles at a defined position with a monoenergetic beam (Fig. 8(b)). The target volume is segmented in several layers, called iso-energy slices (Fig. 8(c)), orthogonal to the beam direction, each corresponding to a different water equivalent beam penetration depth obtained by modifying the energy of the beam (directly through the accelerator for synchrotrons or through an energy-selection system for cyclotrons). The resulting dose delivery technique is a kind of 3D scanning of the target volume.

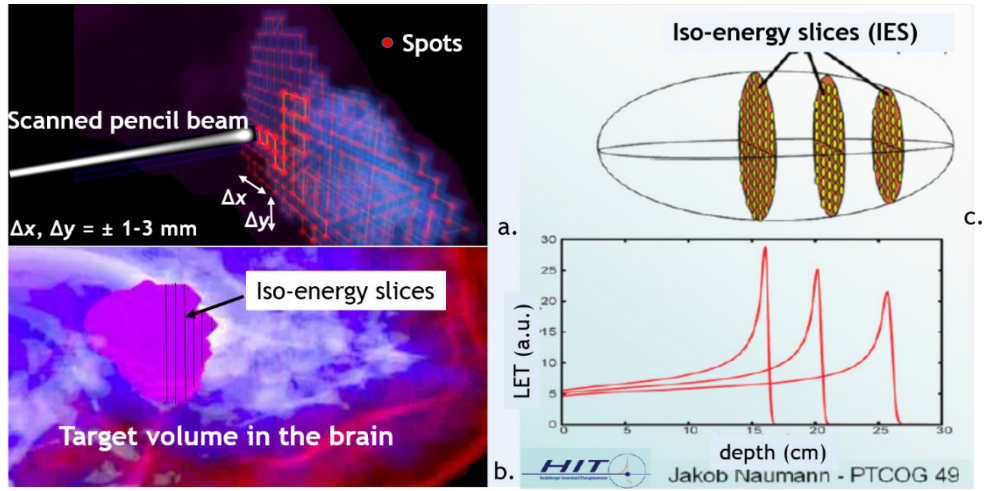

**Fig. 8:** Example of spot distribution (a) for a simulated target volume in the brain (b). (c) Iso-energy slices treated with different Bragg peaks (reproduced from Jakob Naumann presentation at PTCOG49).

The sweeping across the transverse plane is achieved by means of two independent scanning dipoles for horizontal and vertical beam deflections located at the end of the extraction line and several metres upstream of the patient. By changing the energy step and performing the irradiation slice by slice, a tumour of arbitrary shape can be successively irradiated from its most proximal to its most distal part or vice versa. This irradiation technique, only possible with charged particles, allows reaching very high dose conformation in radiotherapy, minimizing the radiation burden on healthy tissues. It reduces the neutron dose to the patient and removes the need for patient-specific devices. In addition, the dose modulation inside the target volume facilitates the application of individualized treatments with local boosts or other desired non-homogeneous dose distributions.

A careful treatment planning assigns the energy, number of particles and transverse position to each spot in order to deliver the optimum dose distribution to the patient. Multiple radiation fields, each characterized by a fixed beam entrance direction, are aimed by using a rotating gantry or, when no gantry is available, by rotating the patient. An isocentric plane is defined for each field, perpendicular to the beam entrance direction, which contains the axis of rotation, the isocentre being the origin of its reference system. Spot transverse positions are typically projected to the isocentric plane and specified relative to its reference system.

To obtain a range of more than 300 mm, the maximum beam energy is about 250 and 430 MeV/u for protons and carbon ions, respectively. The required field transverse size typically ranges from $200 \times 200$ mm² to $300 \times 300$ mm², with a 150 mm length for the longitudinal direction. These cover most of the targets that are usually treated. The required overall precision of the spot position is about 0.5 mm.

### 2.2.1 *Flavours of scanning*

As shown in Fig. 8(c), in pencil beam scanning the dose is generally painted with the pencil beams grouped in discrete energy layers (all beams of the same energy are sequentially applied and only then is the beam energy changed). There are several options to deliver the dose of an energy layer. The choice among them can influence the precision and the treatment dead time and provide different repainting capabilities (by repainting capability, we mean the possible number of repaintings used to reduce the error caused by target motion) [7-9].

Three methods of painting energy layers are available: (a) discrete spot scanning, (b) raster or quasi-discrete pencil beam scanning and (c) continuous line scanning.

With *discrete spot scanning* each pencil beam is delivered statically: after irradiation of a spot the beam is switched off and the position of the beam is changed to the next spot application. Once the dipoles have reached the new designated values, the beam is turned on again. This technique requires a fast beam shut-off system that can be operated at high frequencies. The dead time between two spots depends on the distance between two spots, on the average scanning magnet velocity and on the set-up time including all electronic delays. It is used at PSI [10], at MD Anderson [11] provided by Hitachi Co. and at the Rinecker Proton Therapy Center provided by Varian Co. At PSI Gantry 1, instead of moving the beam by scanning magnets in both transversal directions, it is also possible to move the patient table. As table motion is slow compared to scanning the beam transversally or even to changing the beam energy using the layer stacking method, the table movement is chosen as the slowest varying scan direction [10]. This is in contrast to most other facilities, where changing the beam energy takes most time and where the longitudinal scan direction is the slowest.

The pencil beam *raster scanning* (or *quasi-discrete scanning*) technique is a type of spot scanning that does not turn off the beam between two spots if these are close enough. This dose delivery method requires fast scanning magnets to minimize the dose between two points. This method was developed at GSI [12] and then implemented in the new synchrotron-based facilities like Heidelberg Ion-Beam Therapy Center (HIT) [13], Centro Nazionale di Adroterapia Oncologica (CNAO) [14], Austrian Hadron Therapy Centre (MedAustron) [15] and National Institute of Radiological Sciences (NIRS) [16, 17].

The *continuous line scanning* method has been implemented on Gantry 2 at PSI [18]. The delivery of each spot is replaced by the corresponding dose segment, a line piece that connects spots together over ±half of the grid distance between spots. A line segment is delivered with constant (highest possible) velocity and the required dose rate (varying spot dose) is controlled by adjusting the beam intensity at the ion source. By keeping the beam moving continuously at highest speed on a continuous path, the repainting capability is maximized and the errors due to organ motion are reduced. The required dose delivered along a line piece can be adjusted by changing either the beam intensity or the sweeper velocity.

The instrumentation required to perform such precise dose deliveries are: (a) two dedicated fast scanning magnets, one steering the beam along $Y$ and the other along $X$. These magnets have to be operated with advanced power supplies. (b) The online beam monitors and (c) a dedicated real-time control system. In the following we will discuss in general the specifications giving more detailed descriptions of the solutions adopted at CNAO [14].

### 2.2.2 *The scanning system*

The matching of the dose profile to the tumour volume is obtained with the superposition of spots aimed to a specific position by changing the current circulating in the scanning dipoles. This process is supervised by a delivery control system which uses beam monitors for the online measurement of the beam position and fluence (see Section 3); when the prescribed number of particles in a spot is achieved the system steers the beam to the next spot.

The scanning system must be as fast as possible especially when the beam is not switched off during the movement between spots belonging to the same slice. For example, at CNAO, the fluence delivered during the transition from a spot to the next is ascribed to the destination spot [19]. The transition time, not accounted for by the treatment planning computation, influences the distribution of the delivered dose and has to be minimal [7, 8]. Additionally, with fast scanning capabilities, the overall treatment time is reduced compared to slower systems and the repainting option to treat moving targets can be implemented [9]. In the case of high repainting, one has to consider that the transient dose between spots becomes very significant compared to the static part of the spots.

In order to limit the dose inhomogeneity caused by the beam movement between spots within a few percent (±2.5% as clinically required), it has been estimated that, for typical beam fluxes of ~$10^9$ p/s and ~$10^8$ $C^{6+}$/s, a beam scanning velocity in excess of 20 mm/ms should be achieved. Considering a spot spacing of 1 to 3 mm, this implies a transient time lower than 200 μs for all the beam rigidities. For example, the Heavy-Ion Medical Accelerator in Chiba (HIMAC) is equipped with a fast scanning system [20], which provides beam-scanning velocities of 100 and 50 mm/ms for horizontal and vertical beam movements when measured at the isocentre. It allows covering a uniform 2D field having a 100 × 100 mm² size and spot spacing of 3 mm in a time as short as 40 ms. To fulfil these requirements, strong efforts are in general devoted to develop the fast scanning magnet and its power supply, the high-speed control system and the beam monitoring [20].

The scanning elements, dipole magnets and power supplies, designed for both protons and carbon ion beams, are the most demanding because of the wide beam rigidity ($B\rho$) range, between 1.1 T m for protons at 60 MeV/u and 6.3 T m for carbon ions at 400 MeV/u. For such elements, high ramp speed, low hysteresis and good accuracy are key points in the design.

As an example, a fixed horizontal beam line is sketched in Fig. 9, where the two dipoles are at distances of 5 and 6 m upstream of the isocentre. To cover with pencil beams a surface of 200 × 200 mm², a maximum bending angle of 16 mrad is necessary.

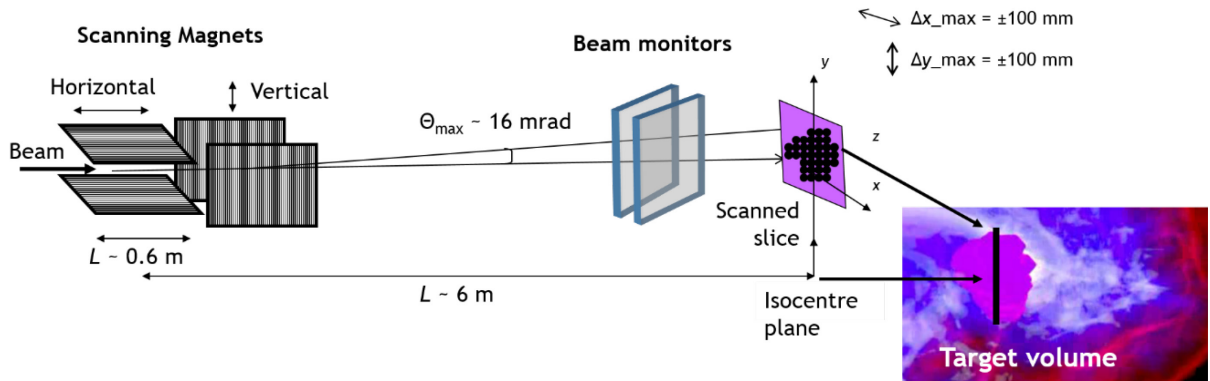

**Fig. 9:** Example of a fixed horizontal beamline for modulated spot scanning delivery

For the vertical line at CNAO, larger bending angles of 21.5 mrad are required because the scanning dipoles are closer to the isocentre due to the additional 90° bending magnet. At the boundaries the operating points of the scanning system of the CNAO treatment lines are listed in Table 2.

**Table 2:** Extreme operating points of the scanning system of the CNAO horizontal (above) and vertical (below) treatment lines. For each beam rigidity the maximum magnetic field and the maximum intensity of the power supplies for covering the 20 × 20 cm² field are reported.

| H line | | | | | | |
|---|---|---|---|---|---|---|
| **Particles** | **Energy [MeV/u]** | **(Bρ) [T m]** | **$B_x$ [mT]** | **$B_y$ [mT]** | **±$I_x$ [A] or [mA/0.1 mm]** | **±$I_y$ [A] or [mA/0.1 mm]** |
| Protons | 60 | 1.14 | 30.47 | 33.99 | 59.05 | 65.87 |
| Protons | 250 | 2.43 | 64.95 | 72.46 | 125.87 | 140.43 |
| C ions | 120 | 3.26 | 87.13 | 97.21 | 168.86 | 188.39 |
| C ions | 400 | 6.36 | 169.98 | 189.64 | 329.42 | 367.52 |
| **V line** | | | | | | |
| Protons | 60 | 1.14 | 35.04 | 44.32 | 67.91 | 85.89 |
| Protons | 250 | 2.43 | 74.70 | 94.48 | 144.77 | 183.10 |
| C ions | 120 | 3.26 | 100.22 | 126.74 | 194.23 | 245.62 |
| C ions | 400 | 6.36 | 195.51 | 247.27 | 378.9 | 479.21 |

At the isocentre plane the beam displacements $\Delta x$ and $\Delta y$ are given by $\Delta x = \alpha_x\, D_x$ and $\Delta y = \alpha_y\, D_y$ [m], where $D_x$ and $D_y$ are the distances between the scanning magnets ($X$ and $Y$) and the isocentre. Assuming small beam deflection angles, $\alpha_x$ and $\alpha_y$, the relation between $\alpha$ and $B$ is the following:

$$\alpha = \frac{\int B dl}{B\rho} \quad [\text{rad}] \tag{1}$$

where $B\rho$ is the magnetic rigidity and is equal to $p/q$, $p$ being the beam momentum and $q$ the charge.

In a first approximation the relationship between the variation of magnetic field and beam displacement is given by

$$\Delta B = \frac{\Delta x \left(B\rho\right)}{D_x l_m} \tag{2}$$

when assuming a constant field in the magnet length $l_m$.

Finally, the relationship between magnet current ($I$) and magnetic induction ($B$) in the dipole gap of height $h_{gap}$ is given in first approximation by (3)

$$B = \frac{NI\mu_0}{h_{gap}}. \tag{3}$$

For constant $B$, if $\int B dl$ is the magnetic length ($l_m$) of the magnet, $h_{gap}$ the dipole height and $D_x$ and $D_y$ the distances from the isocentre, the beam steps $\Delta x$ and $\Delta y$ are given by Eqs. (4) and (5):

$$\Delta x = \Delta I \left[ \left(N\mu_0 / h_{gap}\right) / \left(B\rho\right) \right] D_x l_m \ , \tag{4}$$

$$\Delta y = \Delta I \left[ \left(N\mu_0 / h_{gap}\right) / \left(B\rho\right) \right] D_y l_m . \tag{5}$$

In summary, several parameters influence the design and the performance of the scanning systems:

–  Beam rigidity ($B\rho$ characterizes the field strength required to bend the beam);
–  Scanning speed (sets the maximum transit time between spots);
–  Distance between spots (minimum and maximum beam shifts);
–  Maximum field dimension (characterizes the maximum bending angles required);

- Scanning dipole positions (with $B\rho$ and field dimension determine the maximum bending angles);
- Magnetic length ($l_m$) of the dipole;
- Current ramp rate (indicating the required speed d$I$/d$t$ of the power supplies);
- Beam intensity (determines the minimum spot duration);
- Dose modulation (fluence per spot proportional to the spot duration, which also sets the time between two different current settings).

As examples, the NIRS scanning system at HIMAC is described in Section 2.2.2 while the CNAO scanning dipoles and power supplies are reported in Sections 2.2.3 and 2.2.4.

### 2.2.3    The NIRS scanning system at HIMAC [20]

The new treatment facility at HIMAC is equipped with a 3D irradiation system for pencil beam scanning. The basic parameters of the scanning system are described as follows. To obtain the range of more than 300 mm, the maximum energy is chosen as 430 MeV/u. The required field size is 220 × 220 mm² for the transverse directions with a 150 mm length for the longitudinal direction. This covers most of the targets. Under these conditions, the new system must be as fast as possible to treat the moving target with rescanning. Based on the conceptual design study, the system was designed to provide a modulated dose delivery with beam scanning velocities of 100 and 50 mm/ms at the isocentre. These scanning velocities enable us to achieve the fastest irradiation time of around 40 ms for an example uniform 2D field having a 102 × 102 mm² size with spot spacing of 3 mm. The fast scanning magnet and its power supply, the high-speed control system and the beam monitoring were developed to fulfil these requirements.

The beam line, shown in Fig. 10, consists of the two scanning magnets (SMX and SMY), two screen monitors (SCN1 and SCN2), main and subflux monitors (DSNM and DSNS), position monitor (PSN), mini ridge filter (RGF) and range shifter (RSF). To achieve the fast beam scanning at the isocentre, the distances from SMX and SMY to the isocentre are designed to be 8.4 and 7.6 m, respectively. The vacuum beam exit window is made of 0.1 mm thick Kapton and located 1.3 m upstream from the isocentre. Beam monitors, RGF and RSF, are installed downstream of the vacuum window. The primary beam shutter (FST) and the neutron shutter (NST) will be placed just downstream from the SCN2 indicated by the dotted arrow in Fig. 10.

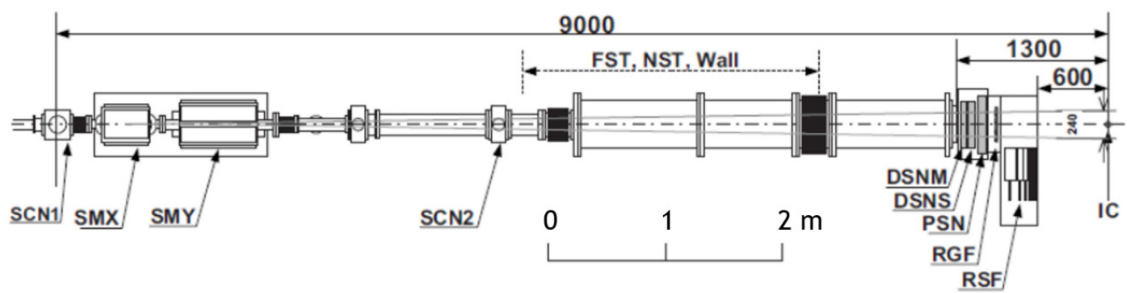

**Fig. 10:** Layout of the HIMAC scanning-irradiation system. SMX, horizontal scanning magnet; SMY, vertical scanning magnet; SCN1 and SCN2, screen monitors; DSNM and DSNS, main and subflux monitors; RGF, bar ridge filter; PSN, beam position monitor; RSF, range shifter; (FST), primary beam shutter; NST (neutron shutter). Unit: mm.

### 2.2.4    The CNAO scanning dipoles

The design of the scanning system had to consider several important and somewhat conflicting features: a good field uniformity, a limited hysteresis, a good linearity up to the largest magnetic fields and a good sensitivity, the latter being required for achieving a good scanning precision when low beam rigidities are used. In addition, a large field ramp was required in order to reach the desired ramp speed of 62 T/s to guarantee a scanning speed larger than 20 m/s for the all beam rigidities [21].

These features were met at CNAO by using for the two identical magnets lamination material with high saturation induction, low and uniform coercivity, and steel with a minimum amount of impurities and large grain size. Very thin (0.35 mm) yoke laminations were glued together to improve the torsion stiffness during the magnet assembly. The maximum magnetic field of 0.31 T is reached with a current of 606 A circulating in the coils. The magnet length is 553 mm and the gap size is $130 \times 140$ mm$^2$ with an inner good-field region of $120 \times 120$ mm$^2$. The coil is divided into three subcoils to optimize the field homogeneity, which was measured to be better than 0.2% in the good-field region. Several aspects affect the behaviour of the magnets when operated at high frequency.

First, the eddy currents induced in the magnet components (the conductor, the iron lamination and the mechanical structure) cause the field in the gap to be attenuated by a quantity $\Delta B(t)$ when a field ramp is applied. The value of $\Delta B$ depends on the imposed field ramp and on the geometrical and electrical characteristics of the magnet components. A careful design of the CNAO magnets limits the deviation to 1 G, decaying with a time constant of 300 µs due to eddy currents in the iron lamination with a field ramp of 62 T/s. A similar effect due to the eddy currents in the magnet conductor induces decays with a longer time constant (3 ms) but with much lower deviation (0.3 G).

Second, because the magnet inductance is a rate-dependent impedance, it has to be considered. For example, the CNAO magnet has an inductance of 4.4 mH and the electrical resistance is 26 m$\Omega$. The load of the dipole magnet can be approximated at first order to a $R$–$L$ circuit with a time constant of 170 ms. This value is three orders of magnitude larger compared to the average time interval required to move the beam between two spots; a dedicated and advanced power supply (PS), described in the next section, was developed to overcome this limit.

A picture of the CNAO scanning dipoles placed on one of the three horizontal beam lines is shown in Fig. 11.

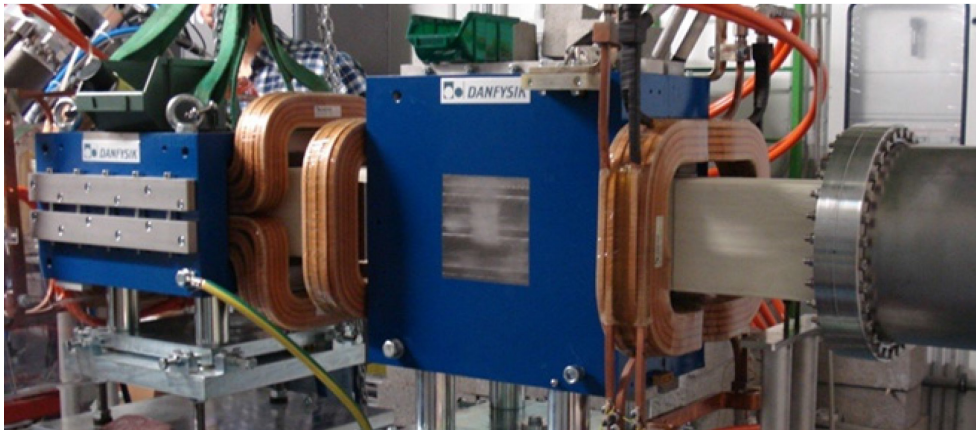

**Fig. 11:** Picture of the CNAO scanning dipoles for the horizontal beamline

### 2.2.5 The CNAO power supplies for scanning magnets

The time constant of scanning dipoles (higher than 150 ms) has to be shortened by three orders of magnitude to reach the required scanning speed of 20 mm/ms. The mechanism implemented for the CNAO PS is based on the pre-emphasis concept by delivering a large voltage step which is then aborted when the current is close to the required value. The precise adjustment is achieved then via smaller voltage steps.

The power supply is composed of three main modules, sketched in Fig. 12: a booster (BO) and two active filters (AFs). The BO is a high-voltage–high-current insulated gate bipolar transistor (IGBT) H-Bridge whereas the AFs are IGBT H-Bridges used in interleaved modulation.

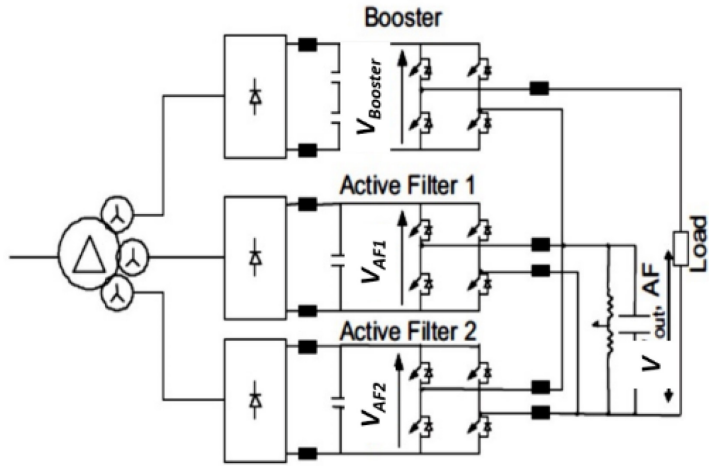

**Fig. 12:** Schematics of the CNAO power supply for the scanning dipoles

The BO provides a large voltage step (±660 V) with a ramp speed exceeding 100 kA/s when a current step larger than 2.5 A is demanded. The BO is switched off by a comparator when the current reaches a value close to the set value within ±0.6 A. The current is then driven to the final value by the control loops of the AFs. For current steps below 2.5 A only the AFs are used and the BO is short circuited.

The AFs provide both precise voltage adjustments during the transient and control for steady-state current. The precision of the delivered current is 55 mA, corresponding to 100 ppm for $I_{max} = 550$ A. The AF control loop sensitivity is 36.4 mA.

The connection between the power supply and the magnet is achieved through a 10 m long shielded cable.

### 2.3 Dose delivery constraints for different particles and accelerators

Depending on the particle species used in particle therapy, different constraints apply to the dose delivery system and instrumentation. We will limit the discussion to protons and carbon ions, which are presently the only species used in clinics, and to the two types of accelerators available for hadron therapy, the cyclotron and the synchrotron. We will see how different beam characteristics lead to some basic differences in the dose delivery concept and instrumentation.

Briefly, the cyclotrons provide to the DD system only proton beams with a single fixed energy and with continuous and steady current. Therefore, to achieve a longitudinal dose distribution requires a fast degrader in the accelerator or range shifter plates for energy degradation in the dose delivery. Because of the presence of the absorbers, the beam flux produced by the accelerator must be higher compared with the clinical flux and the activation of the material along the beam line and in the dose delivery instrumentation must be carefully considered.

The synchrotrons accelerate both protons and heavier ions at fixed energies which can be selected within a large set of values. The beam energy range is usually chosen to avoid the use of range shifters. The synchrotrons work in pulsed mode, i.e. the extracted beam is not continuous and, during the extraction, called spill time, the beam current is generally not constant. The beam energy can be changed spill by spill through the real-time dose delivery control.

The transversal beam spread does not depend on the accelerator, so scattering and scanning modalities can be used both with cyclotrons and with synchrotrons.

The choice of spreading the beam with the passive scattering techniques allows delivering a dose with protons as was done with photons: through large and uniform radiation fields entering in the accelerator nozzle where patient-specific static devices conform the dose distribution. In these systems, the goal of the dose delivery is first to spread the protons in order to create a large square or circular field and then to degrade and absorb particles to conform to the required field shape. The robust and safe approach of passive scattering systems led to a large number of proton facilities. The technology for proton medical accelerators and for the dose delivery instrumentation has been widely improved in the last decade resulting in solutions available on the market with affordable costs. Additionally, in the treatment room gantries are widely adopted allowing patient irradiation from any beam direction.

The carbon ions show better physical properties providing a better conformation to the target when the modulated scanning technique is used and are particularly effective for some radio-resistant tumours. However, the low scattering angle and the projectile fragmentation is a constraint to use these particles with the passive scattering beam delivery techniques. As a consequence, the treatments with carbon ions have started later, in Europe at GSI first in the 1990s, where the active scanning technique to spread laterally the beam has been first designed and developed [12]. The main disadvantage of carbon ions is that a cyclotron accelerator for therapeutic heavy ions is still not available and presently only large synchrotrons can be used, increasing the system complexity and cost. Moreover, the beam particles fragment in the patient body, creating lighter particles with larger penetration depth, generate a tail beyond the Bragg peak, giving an unwanted dose after the treatment volume. Beam particles and fragments have different and variable relative biological effectiveness (RBE); this is why the radio-biological dose evaluation is complex and many details are still under investigation. Moreover, for carbon ion beams the gantry system is very large and expensive, so, up to now, a single system is clinically used [22], while the other carbon ion facilities have only fixed beam lines.

## 3 The beam monitoring system

### 3.1 Introduction

The delivery of the dose has to be performed to ensure the safety of the patient. Therefore, accurately monitoring the delivered dose in particle therapy is mandatory and is equivalent to accurately monitoring the correct delivery of the number of ions of the primary beam.

Additional checks may be required, which consist in measurements to control the beam parameters, listed in Table 1.

The treatment prescription is provided as a DICOM RT file by a TPS and consists of a set of data to set up the accelerator, the beam line and the beam shaper or scanning devices before and during the treatment delivery.

Passive scattering systems have different constraints and requirements for the beam monitors compared to systems for modulated pencil beam scanning DD systems. The latter must change the beam settings during the treatment; so, faster and real-time controls are mandatory.

For the scattering delivery technique, the type and position of scatterers, range shifters and compensators are provided by the TPS and verified before the irradiation starts. The total amount of monitor units, each corresponding to a given amount of dose, is also provided among prescriptions and used by the intensity beam monitor to stop the treatment when the total desired dose has been achieved (i.e. at the end of the field irradiated).

For pencil beam scanning deliveries, the prescriptions consist of data to define pre-treatment accelerator and beam line settings followed by a list of spots defined in terms of number of particles (or monitor units), energy and position. These quantities are used online by the dose delivery instrumentation (i.e. beam monitors and scanning devices) to guide the treatment. Note that additional beam monitoring requirements exist for pencil beam scanning because of the need to drive the delivery progress by acting on scanning magnets and on beam stopper devices.

The sensitivity of the monitoring system should match the maximum tolerable uncertainties that, for the pencil beam scanning, are listed in Table 3.

**Table 3:** Maximum uncertainties for the beam parameter. Note that the beam energy is not measured online by beam monitors but verified before the treatment starts by checks of the accelerator (for synchrotron) or beam line settings (for cyclotron).

| Beam property | Maximum uncertainty |
|---|---|
| Beam flux | 1–2% of the integral flux |
| Transversal position | 0.5 mm |
| Transversal shape (FWHM) | 1 mm |
| Energy | 1–2% |

### 3.2 Beam monitoring for scattering systems

For the scattering dose delivery technique the required beam characteristics are set before the irradiation and remain the same for the whole irradiation field. The beam fluence, position, shape and symmetry are continuously checked for safety purposes at a low frequency (Hz). However, beam parameter fluctuations within a small range are tolerable because the beam is then scattered over the total target volume. Thus, if the mechanical parts like scatterers, range shifters and compensators are in the correct position and the primary beam has the right energy, the correct beam delivery to the target volume is safe and ensured. The measured beam fluence is usually used by the DD system to stop the treatment when the total desired dose has been achieved.

The beam monitoring in passive beam delivery is generally performed by two independent dose monitoring devices located before the final collimator. One detector works as the main monitor (master) and the second works as the auxiliary (or sub or slave) monitor for redundancy. Additional beam profile and reference dose monitors measure the beam shape and intensity at the exit of the vacuum chamber to verify the particle distribution before the scattering and modulation.

The MD Anderson (Houston, Texas) passive scattering nozzle (Fig. 13) is taken as an example of a dose delivery system with gantry for a proton beam line [23].

After the beam exits the beam transport system, it passes through a vacuum window into the treatment delivery nozzle. As the beam traverses the nozzle it intercepts the following devices: (1) beam profile monitor, (2) reference dose monitor, (3) first scatterer with a range modulation wheel (which form a single, integrated unit), (4) second scatterer, (5) binary range shifter, (6) secondary dose monitor, (7) primary dose monitor, (8) multilayer Faraday cup (MLFC), (9) treatment field aperture, range compensator. The beam profile monitor, reference dose monitor, secondary dose monitor, primary dose monitor and MLFC are used to monitor various aspects of the beam while the remaining devices are used to shape or modify the beam.

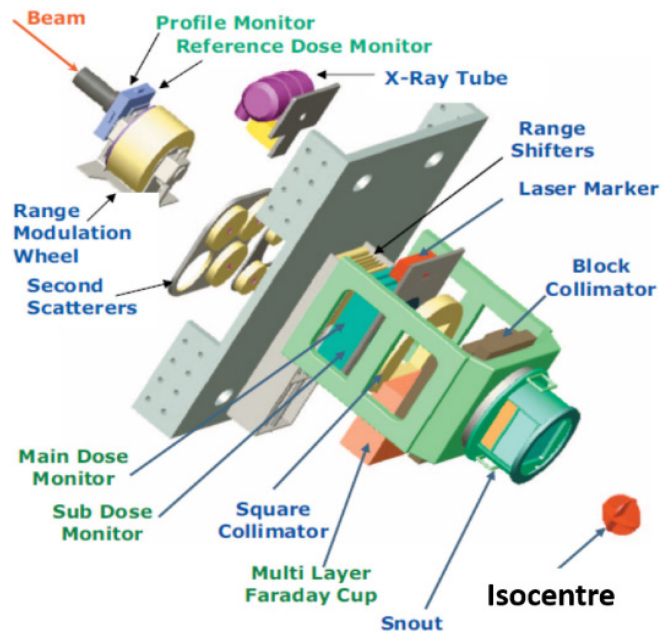

**Fig. 13:** Three-dimensional rendering of the MD Anderson passive scattering nozzle

Another example is the Catana (Centro di AdroTerapia e Applicazioni Nucleari Avanzate) fixed horizontal proton beam line for ocular treatments shown in Fig. 14, where two transmission ionization chambers for fluence measurement are followed by a strip chamber [24] for beam position and shape check.

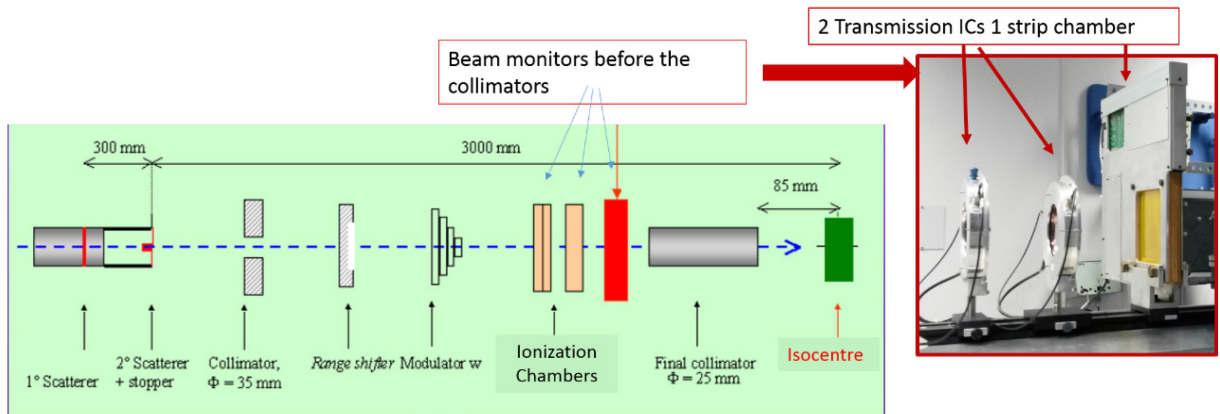

**Fig. 14:** Scheme of the Catana beam line with a picture of the two transmission ionization chambers and the strip chamber (in red).

### 3.3 Beam monitoring for pencil beam scanning techniques

Active scanning techniques require a precise, quick, stable and reliable beam monitoring system: the monitoring has to be repeated for each spot to meet the prescriptions, ensuring in addition the correct position of the spots. When the beam monitor has collected the prescribed amount of charge for a spot, the beam is steered to the next spot position by the DD control through the scanning magnets by changing the currents of the power supplies, see Fig. 15. This procedure is repeated until the last spot of the slice, i.e. a sequence of spots with the same energy, has been irradiated.

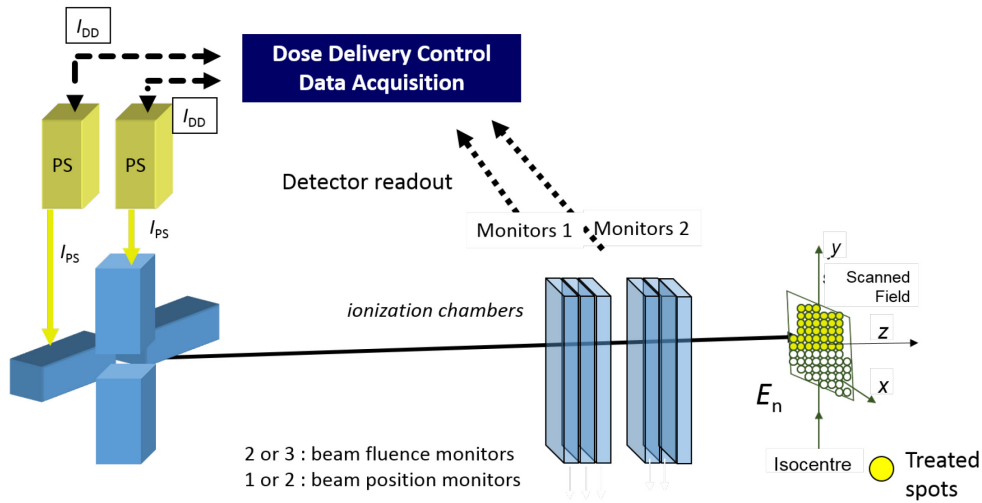

**Fig. 15:** Scheme of a primary beam monitor system for pencil beam scanning

Real-time and fast control systems are required to react to any condition leading to a potential hazard. Moreover, online feedback corrections (for example small corrections of beam position deviations based on beam monitor data) are recommended.

As beam monitors, thin detectors, fast and reliable for measuring beam fluence and position, are located just in front of the patient (Fig. 15). Arrays of parallel-plate ionization chambers with either a single large electrode or electrodes segmented in strips or pixels (see Section 3.1) as well as multiwire chambers (see Section 3.5) are well suited for this purpose. Special care is required in the detector design for high-intensity and pulsed beams. More details will be discussed in the following sections.

Secondary emission monitors (SEMs) may be used only with high dose rates because of their lower sensitivity (see Section 3.6).

In summary, the beam monitors, positioned as close as possible to the patient, have the tasks of measuring:
–   the number of particles (i.e. integrated beam flux, often expressed in terms of monitor units);
–   the transversal beam positions;
–   the transversal beam shape (i.e. FWHM and symmetry).

The beam energy is typically not measured by the online beam monitors but guaranteed through proper checks of the accelerator and beam line settings (for synchrotrons) or by pre-treatment range measurements (for cyclotrons and synchrocyclotrons).

In pencil beam scanning systems that do not stop the beam during the transit between spots, a tiny fraction of the dose is delivered along the path. Since the transit time is typically very short, hundreds of microseconds, the beam monitoring system is unable to measure where the particles were actually delivered. The easiest choice is either to consider these particles as delivered entirely to the spot whose irradiation is terminated or to assign it to the following spot which is being irradiated. The latter is the choice implemented by the DD of CNAO. This choice overestimates the fluence as the beam approaches a given spot and, of course, it underestimates the dose as the beam moves away from the same spot. The two effects partially compensate, the net effect being sizeable for mainly the first and the last spots of a sequence. Moreover, to limit the transient dose when the distance between spot centres is larger than a pre-defined threshold, the irradiation can be paused. Optimization algorithms were also proposed to optimize the scanning path in order to minimize the dose delivered during the transit [7, 8].

### 3.3.1 Ionization chambers

The most widely used type of beam monitor for charged particle therapy is the parallel-plate ionization chamber. Compared to other detectors, it offers several advantages in terms of robustness, uniformity, easiness of operation and minimal perturbation of the beam. A sketch of an ionization chamber is shown in Fig. 16. Two parallel plates, made of metallized foils kept at a constant high voltage difference, generate a uniform electric field. Particles traversing the chamber ionize the gas in the volume bounded by the plates. The thickness of the plates and the gas are chosen to minimize the induced scattering and fragmentation of the beam particles.

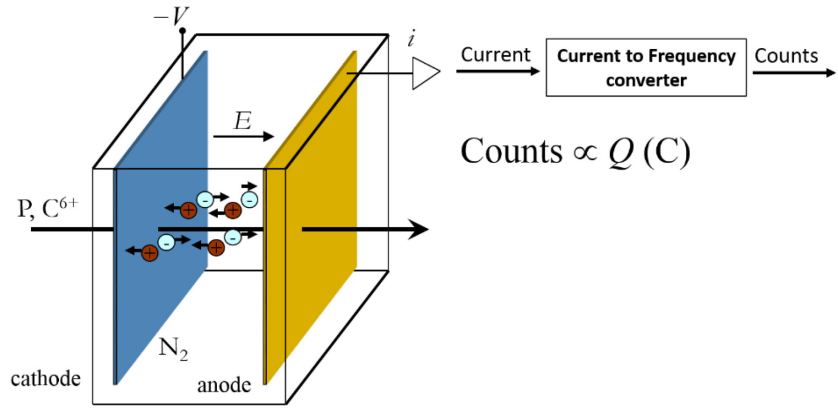

**Fig. 16**: Parallel-plate ionization chamber filled with nitrogen

Ionization chambers work efficiently if a large number of charge pairs are created and collected at the electrodes before they recombine to form neutral molecules. Therefore, recombination of the ionization charges at the highest expected dose rates must be taken into account in choosing

– the type of gas (usually air or nitrogen);
– the plate separation and the operational high voltage (i.e. the electric field, typically higher than 500 V/mm to ensure rapid charge collection with minimum recombination effects).

Also, multiplication of the produced charge should be avoided because it can lead to unacceptable errors in the beam intensity measurement.

There are different mechanisms through which these interactions can take place. Some of these interaction mechanisms can be predicted with a good level of accuracy by using statistical quantities such as cross-section and stopping power. A quantity that is extremely useful for radiation detection is the average energy needed to create an electron−ion pair in a gas. Ions can be formed either by direct interaction with the incident particle or through a secondary process in which some of the particle energy is first transferred to an energetic electron or delta ray. This quantity has to be considered as an effective energy, larger than the gas molecule ionization energy, to account for all the energy loss mechanisms not leading to ionization in the gas.

When the change in particle energy during the detector traversal can be neglected, the average charge produced per particle within the gas gap is given by Eq. (6):

$$\frac{Q}{\int_0^{E\max} f(E)\,dE} = \frac{e\int_0^{E\max} \frac{\left(\frac{S(E)}{\rho}\right)_g \cdot \rho \cdot x}{W(E)} f(E)\,dE}{\int_0^{E\max} f(E)\,dE},$$

(6)

where $(S(E)/\rho)_g$ is the mass electronic stopping power for the gas g at particle energy $E$, $\rho$ is the density, $x$ is the mean free path in the gas, $W(E)$ is the mean energy needed to produce an ion pair for a charged particle of energy $E$ losing $(S(E_p)_g \cdot \rho \cdot x)$ energy in the gas and $f(E)$ is the number of beam particles crossing the chamber with energies between $E$ and $E + dE$. The formula can be simplified if a monoenergetic beam of energy $E_p$ is considered:

$$\frac{Q}{N_p} = \frac{\left(\dfrac{S\left(E_p\right)}{\rho}\right)_g \cdot \rho \cdot x}{W\left(E\right)} \ , \tag{7}$$

where $N_p$ is the total number of particles crossing the chamber. For a monoenergetic beam of uniform intensity the formula is typically written as follows:

$$i = \varepsilon \cdot \Phi \cdot \frac{\left(\dfrac{S\left(E_p\right)}{\rho}\right)_g \cdot \rho \cdot x}{W\left(E\right)} \ , \tag{8}$$

where $i$ is the intensity of the current collected at the electrodes of the chamber, $\Phi = N_p/\Delta t$ is the beam flux and $\varepsilon$ represents the collection efficiency. The charge $Q$ measured at the electrodes is thus directly proportional to the number of particles crossing the gas gap $N_p$.

The practical quantity of interest on radiation detection is the total number of ion pairs created along the beam particle tracks. There is a minimum energy transfer from the incoming particle to the electron necessary so that the ionization of the atom occurs. In most of the gases used for radiation detectors (air, nitrogen, argon etc), the minimum ionization energy is between 10 and 25 eV. However, other mechanisms by which the incident particle may lose energy within the gas do not create ions. Examples are excitation processes in which an electron may be excited to a higher bound state in the atom without being completely removed. Therefore, the average energy loss by the incident particle per ion pair, defined as the W-value, is always substantially larger than the minimum ionization energy. The W-value is found to depend weakly on the particle type and energy and lies within 25–45 eV per charge pair for most gases and types of radiation (Table 4).

**Table 4:** W-value in several gases for electrons (data from ICRU 1979)

| Gas | W-value (eV/i.p.) |
| --- | --- |
| Air | 34.2 |
| He | 41.3 |
| Ne | 35.4 |
| Ar | 26.4 |
| $H_2$ | 36.5 |
| $N_2$ | 34.8 |
| $O_2$ | 30.8 |
| $CO_2$ | 33.0 |

Air- or nitrogen-filled chambers are typical choices and automated corrections for atmospheric pressure and temperature are recommended when the collected charge is transformed to number of beam particles delivered.

The charges created by the incident radiation are called primary charges to distinguish them from the ones produced by ionization caused by primary charge pairs. The W-value represents all such ionizations that occur in the active volume. For a particle that deposits energy $\Delta E$ inside a detector, the W-value can be used to determine the total number of ion pairs produced by

$$N = \frac{\Delta E}{W} \ . \tag{9}$$

In the unlikely case that the incident particle deposits all of its energy inside the detector gas, then $\Delta E$ would simply be the energy $E$ of the particle. However, in a usual case only a negligible part of the total energy is lost in the gas; then the number of ion pairs is a function of the stopping power, as

$$N = \frac{1}{W} \frac{\mathrm{d}E}{\mathrm{d}x} \ \Delta x, \tag{10}$$

where $\Delta x$ is the path covered by the particle. Sometimes it is more convenient, at least for comparison purposes, to calculate the number of ion pairs produced per unit length of the particle track as follows:

$$n = \frac{1}{W} \frac{dE}{dx}. \tag{11}$$

*Drift velocity*

In a gaseous detector, free electrons behave quite differently compared to the ions in the presence of the electric field and therefore the two types of charges should be studied separately. The ions are positively charged and much heavier than electrons and therefore move around quite sluggishly compared to the electrons. In the ionization chambers, the output signal can be measured from the positive or from the negative electrode. In both cases, however, what is measured is actually the charge induced by the change in the electric field inside the active volume. Hence, the drifts of electrons and ions both contribute to the overall output pulse. This implies that understanding the drift of positive charges is as important in a chamber as the electrons.

In the presence of an externally applied electric field, ions move toward the negative electrode with a drift velocity that is much lower than that of electrons. Assuming that $N$ ions are produced in $x = 0$ at $t = 0$, the distribution at time $t$ of these ions can be fairly accurately characterized by a Gaussian distribution of the form

$$dN = \frac{N}{\sqrt{4\pi Dt}} \mathrm{e}^{-(x - v_\mathrm{d})^2 / 4Dt} \ \mathrm{d}x, \tag{12}$$

where $v_\mathrm{d}$ is the drift velocity of ions, i.e. the average velocity of the cloud of ions moving along the electric field lines, and $D$ is a temperature-dependent diffusion coefficient per unit time. The drift velocity is much lower than the instantaneous velocity of ions. Drift velocity is an important parameter, since it indicates how quickly the ions reach the cathode and get collected. It can be fairly accurately predicted from the relation

$$v_\mathrm{d} = \mu^+ \frac{E}{P}. \tag{13}$$

Here $E$ is the applied electric field, $P$ is the pressure of the gas and $\mu^+$ is the mobility of ions in the gas. Mobility is related to the mean free path of the ion in the gas, the energy loss per impact and the energy distribution.

If the chamber is filled with electronegative gases like air, electrons are rapidly attached to the neutral gas molecules giving rise to negative ions drifting with a velocity similar to the positive ions, though in the opposite direction. On the contrary, if a non-electronegative gas like nitrogen is used, free electrons will drift to the anode with a velocity two to three orders of magnitude larger than the positive ions.

Free electrons, owing to their small mass, are rapidly accelerated between collisions and thus gain energy. Along the electric field lines, the electrons drift with velocity $v_d$, which is usually an order of magnitude smaller than the velocity of thermal motion, $v_e$. However, the magnitude of the drift velocity (Fig. 17) depends on the applied electric field $E$ as follows:

$$v_d = \frac{2eEl_{mt}}{3m_e \overline{v_e}},\tag{14}$$

where $l_{mt}$ is the mean momentum transfer path of electrons.

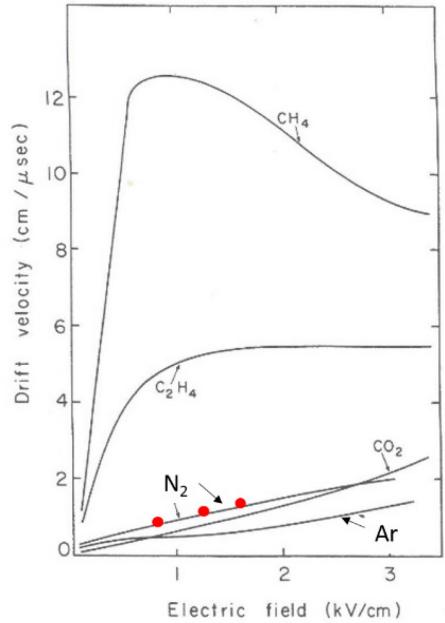

**Fig. 17**: Electron drift velocity as a function of the electric field ($V$/m) for different gases (reproduced from [25])

Table 5 shows the time required to an electron and an ion to cross a 5 mm gap of a nitrogen-filled ionization chamber with the electrodes polarized at 500 V.

**Table 5:** Example of ion and electron collection times for 5 mm gap of a nitrogen-filled ionization chamber

| Charge | Gas | $E$ (V/cm) | $v$ (cm/s) | Gap (mm) | Time ($\mu$s) |
|---|---|---|---|---|---|
| Ions | $N_2$ | 1000 | $3 \times 10^3$ | 5 | 150 |
| Electrons | $N_2$ | 1000 | $10^6$ | 5 | 0.5 |

The current measured by this chamber, given a total charge $Q$ released uniformly in the chamber by the beam passing through, is provided by both ions and electrons collected at the electrodes at different rates as shown in Fig. 18. If the integral of this curve is used to measure the charge $Q$, and hence the number of particles $N_p$ that crossed the chamber, a measurement time long enough to collect all the signal is required or a systematic underestimation will occur. This is particularly relevant for a pulsed beam structure if the number of particles delivered in each pulse needs to be determined with high accuracy.

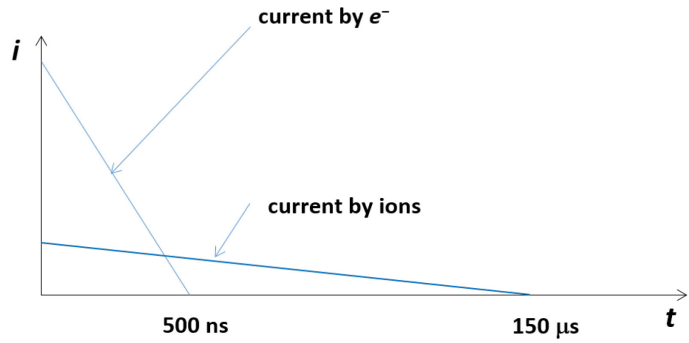

**Fig. 18**: Different contributions to the measured current in a beam monitor characterized by a 5 mm gap filled with nitrogen and supplied with 500 V ($E = 1000$ V/cm).

The charge collection efficiency $\varepsilon$ is mainly determined by the recombination of charges in the gas volume. As the voltage difference between the electrodes increases, the resulting electric field separates the ion pairs with increasing drift velocity, reducing the equilibrium concentration of ions within the gas. The ionization current thus increases due to fewer charges lost to recombination, at first almost linearly with voltage, then more slowly and finally asymptotically approaches the saturation current for the given radiation intensity. At this level, the electric field is large enough to effectively suppress the recombination to a negligible level, and all the original charges created through the ionization process contribute to the ionization current. Increasing the voltage further cannot increase the current because all charges are already collected and their rate of formation is constant. Therefore, the saturation current is the measured current if all the ions formed in the chamber by the radiation are able to reach the electrodes. Different detector designs (i.e. gap and gas type) and settings (i.e. high voltage between electrodes) lead to different saturation curves. In Fig. 19, we show the charge collected as a function of the applied voltage for transmission ionization chambers filled with air with different gaps that have been irradiated with $3 \times 10^7$ carbon ions per spill. For these data, the spill length was about one second.

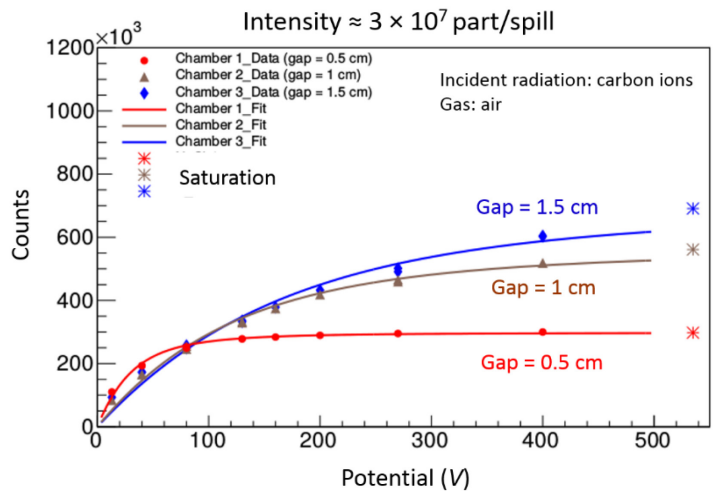

**Fig. 19**: Detector counts (i.e. charge collected) from parallel-plate ionization chambers with different gaps, irradiated with $3 \times 10^7$ carbon ions per spill (i.e. per second), as a function of the voltage.

Parallel-plate geometry simplifies the electrode design generating a uniform electric field. The electrodes of the beam monitors are perpendicular to the particle beam axis and a single large-area electrode is used to measure at high rate, typically larger than 100 kHz, the beam flux. Parallel-plate geometry can also be used efficiently for measuring the beam shape, beam position and beam fluence using segmented electrodes.

In the segmented type, each element measures the collected charge independently of the other elements. Such chambers have a variety of uses such as measuring the beam profile and position or measuring the two-dimensional beam profile distribution. The layouts of such devices depend on the required spatial resolution and on the total area to be covered. As is shown in Fig. 20, strip chambers can provide high resolution (~100 μm) and faster (10 kHz) beam position evaluation compared with pixel chambers (~200 μm and 5 kHz). In Fig. 20, different anodes with different segmentations are shown.

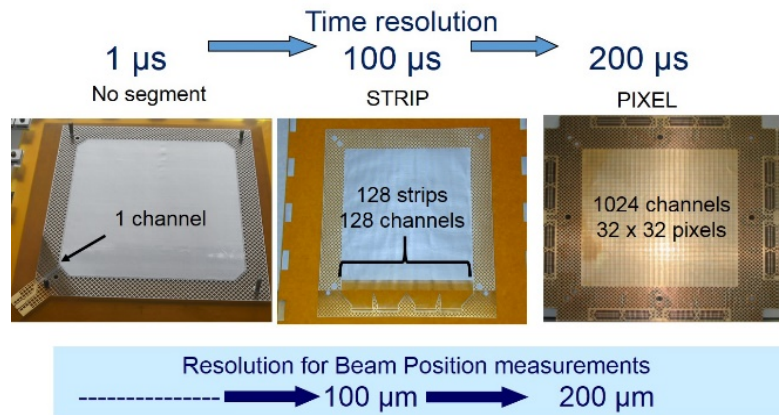

**Fig. 20:** Picture of anodes for ionization chambers segmented in different shapes

The practical limit on the reduction of the size of the collecting elements is determined by the maximum size of the irradiated fields, by the maximum density of channels on the anode (or cathode) surface and by the detector readout capability.

The detector response should be position-independent, especially for pencil beam scanning dose deliveries, because the beam spot is scanned across the whole detector. The beam-induced ionization in the gas detector should be identical at any position in the detector while the beam direction has a normal incidence with the detector. The main parameter affecting the uniformity is the flatness of the electrodes that may determine variation of the gas gap.

The charge collected in the ionization chambers depends on the pressure and temperature of the gas, as well as on the voltage across the gap. These quantities have to be continuously measured and checked; the read values are used to correct the gain of the chambers. Appropriate interlock procedures are required whenever any of the values is outside the expected range.

For passive scattering dose delivery, beam-centring systems (including segmented ionization chambers), which are capable of detecting misalignments between the central axes of the beam and the scattering devices, are required. Dose-monitoring detectors can either intercept the entire beam area or just the central portion. The former requires larger detectors and measurements are more reliable as they are less dependent on beam alignment variations. Beam steering is much more sensitive in double- than in single-scatterer DD systems.

*Beam position and width*

The beam position is usually evaluated online through the measurements of ionization chambers segmented in strips or of multiwire detectors (see Section 3.5). The distributions of the charges during a typical clinical application are shown in Fig. 21 as a function of the position; they represent the projections of the beam phase space (or beam size) respectively on the horizontal ($X$) and vertical ($Y$) directions.

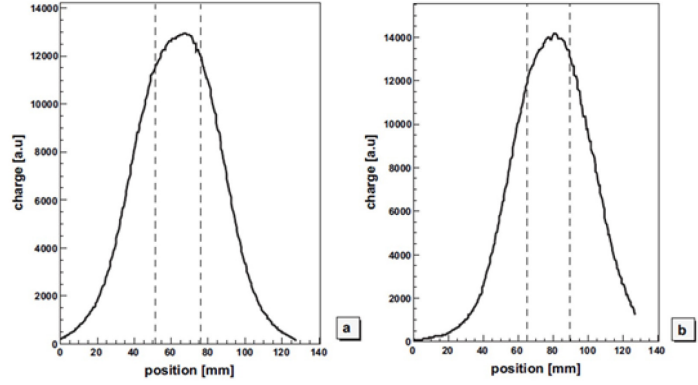

**Fig. 21:** Transversal beam profiles as measured from strip-segmented ionization chamber

An example of a charge distribution as measured with a strip chamber is shown in Fig. 22.

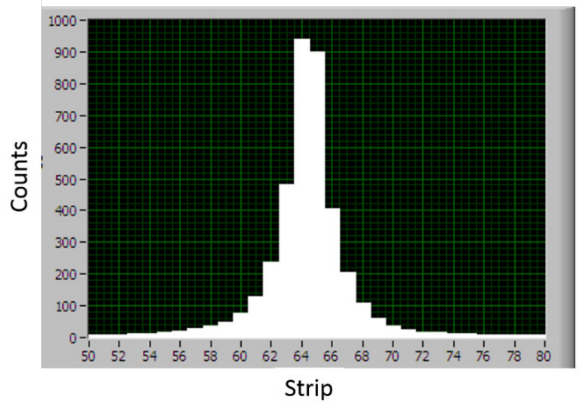

**Fig. 22:** Charge distribution over strips for a proton beam

The beam position can be evaluated as a weighted average through the following equations:

$$X = \frac{\sum x_i^{\text{strip}} \times S_i}{\sum S_i} \; ; \qquad Y = \frac{\sum y_i^{\text{strip}} \times S_i}{\sum S_i}, \tag{15}$$

where $x_i$ and $y_i$ are the coordinates of the strip centre, $S_i$ is the charge signal measured in the strip $i$ and the sum extends over all the strips of the chamber.

To avoid wrong position estimation because of noise or background currents, the values measured by strips far from the beam should be neglected. For example, at CNAO, the strips which measure a single count are not considered in Eq. (15) because this corresponds to the average expected background current for each readout channel.

The FWHM or the σ of the charge distribution is used to estimate the transversal beam width:

$$\text{FWHM}_x = 2.35 \sqrt{\frac{1}{S} \sum_i S_i \left( x_i - X_{\text{beam}} \right)^2}, \tag{16}$$

where $S = \sum_i S_i$ and $X_{\text{beam}}$ is the beam position along $x$ as from Eq. (15).

*Beam shape (flatness and symmetry)*

For passive scattering systems, where a uniform transverse beam distribution is covering the irradiation fields, beam monitors can be used to check the beam flatness and symmetry before and during the treatment. The flatness $Fl$ is used to estimate the maximum percentage of deviation from the average dose delivered in a reference region; following the conventional definition used in radiotherapy, it is given by Eq. (17):

$$Fl = \frac{\left(D_{max} - D_{min}\right)}{\left(D_{max} + D_{min}\right)} \times 100,$$  (17)

where $D_{max}$ and $D_{min}$ are the maximum and minimum doses measured in the selected uniform region; a negative sign is conventionally assigned to $Fl$ if $D_{max}$ occurs on the left-half part of the detector.

The symmetry $S_y$ is defined by considering the integrated absorbed doses $D_l$ and $D_r$ in each half of the field about the centre of the selected region, where the quantity $\left|1 - \frac{D_r}{D_l}\right|$ reaches its maximum:

$$S_y = \left(\frac{D_r}{D_l}\right) \times 100.$$  (18)

The clinical tolerances imposed on these parameters are typically $Fl < 2$–3% and $|S_y - 100\%| < 3\%$.

A symmetry in the beam size is essential, especially at the entrance of passive scattering and gantry systems. In the latter, the shape and position of the beam at the patient and the transmission through the gantry should not depend on the gantry angle.

The beam symmetry can also be evaluated by using the skewnesses $\gamma_x$ and $\gamma_y$ calculated from the transversal beam projections, as follows:

$$\gamma_x = \frac{\sum_i S_i \cdot \left(x_i - \mu_x\right)^3}{\sum_i S_i \cdot \sigma_x^3}$$  (19)

and similarly for $y$. In Eq. (19), $S_i$ is the charge measured in the strip $i$, $x_i$ is the coordinate of the strip centre and $\mu_x$ and $\sigma_x$ are the mean and the r.m.s. of the distribution. For a symmetric beam projection the expected value of the skewness is zero, while positive and negative values indicate asymmetric beam particle distributions as shown in Fig. 23.

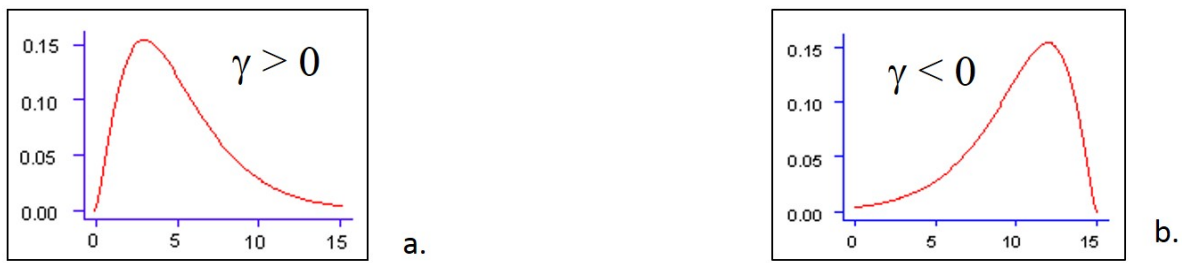

**Fig. 23:** Examples of asymmetric particle beam distributions with positive (a) and negative (b) skewness values

### 3.4    Detector readout

The current collected by each anode of the ionization chambers (i.e. the single channel of the integral beam monitors and each strip or pixel of segmented anodes) is converted into a digital frequency of increments or decrements of a counter, proportional to the current itself. The time integral of the frequency corresponds to a number of counts that is proportional to the collected charge and thus to the energy released in the detector active area. The resulting counts are stored in a register. In the more advanced applications ASICs (application-specific integrated circuits) are specifically developed equipped with ADCs (analogue to digital converters) and registers. Real-time and safe data operations, required for control in clinical applications, are often demanded by field programmable gate arrays (FPGAs), which, besides a high flexibility, guarantee fast and deterministic data processing. A direct connection of the FPGA output with the interlock collector is recommended to perform prompt and safe actions. A scheme with the sequence of signals for a typical detector readout is shown in Fig. 24.

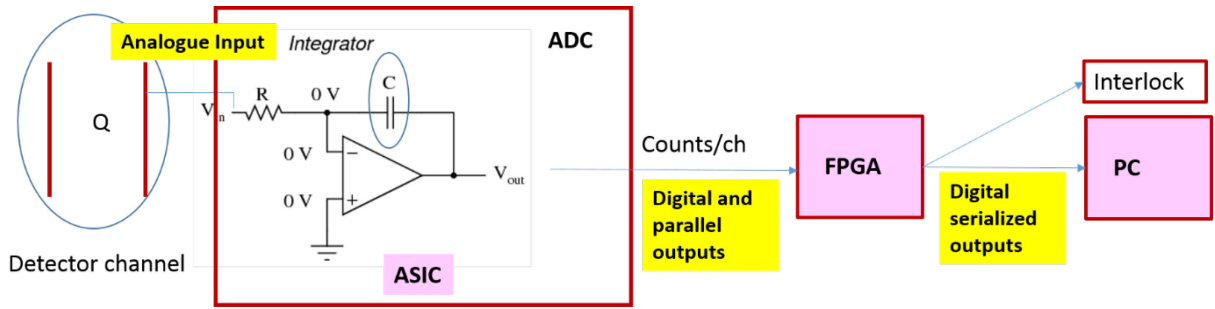

**Fig. 24:** Sequence of signals for a typical detector readout

### 3.5 Multiwire ionization chambers

A multiwire ionization chamber [25], in its simplest form, is a grid of parallel thin anode wires between two cathode planes as shown in Fig. 25(a). Under application of a symmetric voltage difference, ionization electrons released in the upper and lower gas volumes drift with a constant speed to the anodes, where, because of the high field gradient, they are amplified in an avalanche. The backdrift of ions produced in the avalanche away from the anode induces a negative charge on the wire and positive charges in all surrounding electrodes (adjacent anodes and cathode planes).

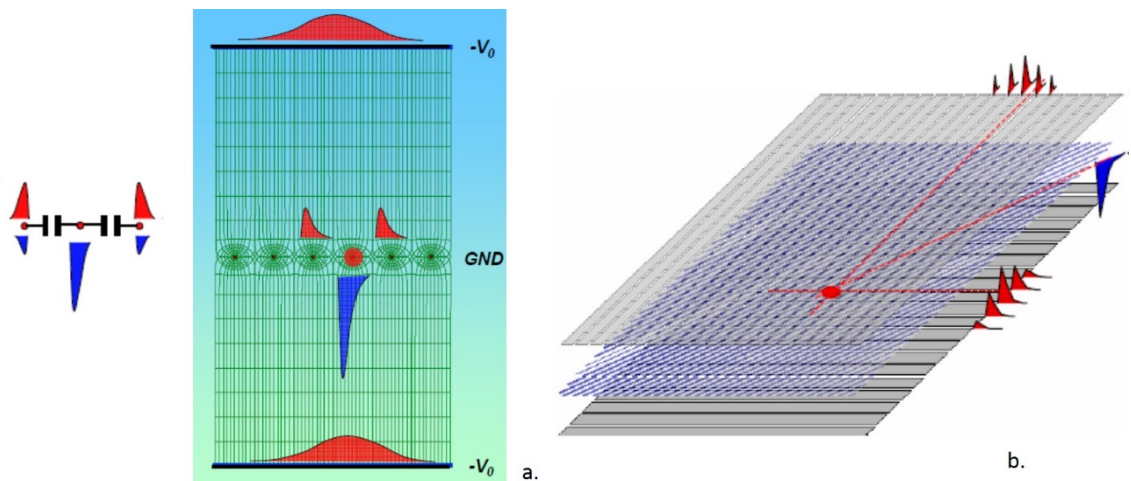

**Fig. 25:** (a) Scheme of a MWIC; (b) perpendicular cathode strips

Signal planes are alternated with high-voltage planes. The high-voltage plane can also be a solid foil conductor instead of parallel wires. While the common application of wire chambers in particle physics is to measure individual particles, in medical applications wire chambers are often used as integrating devices: the signal collected on a wire during a given time is proportional to the number of particles crossing the gas volume around the full length of the wire. In Fig. 25(b), an example of perpendicular cathode strips (printed circuit or wire groups) is shown. The centre of gravity of the induced charge distribution on cathode strips provides the *X* and *Y* projections of the avalanche position. The gain of such devices is dependent on the gas used, its pressure and the wire spacing, and is an exponential function of the high voltage. Clinically, with these devices a submillimetre resolution with a wire pitch of approximately 1 mm can be reached. Wire chambers are typically operated in a proportional mode and called multiwire proportional chambers (MWPCs), where a multiplication of the initial ionization occurs. However, the required high electrical fields can easily lead to saturation for high-intensity beams and only the beam tails can be monitored. This problem can be overcome by operating the wire chambers in an ionization mode, thus improving their performance. Multiwire chambers are primarily used in many centres for beam monitoring during beam line tuning and for on line beam control, for example at the first hadron therapy medical centre (Loma Linda University Medical Center [26]), at the GSI experimental carbon ion therapy beam line and at HIT [13]. The fact that good spatial resolution is obtained with a small number of signals and variable sensitivity makes these devices extremely practical for this purpose.

### 3.6    Secondary emission monitor

A secondary emission monitor (SEM) [27] is particularly well suited for large proton fluences. It consists of one or more thin metallic foils mounted in an ultra-high-vacuum enclosure. As the foil is traversed by protons, electrons are released, resulting in a net current flow that provides the signal. A thin Al foil is placed in the hadron beam path at 45° with respect to the beam direction. The energy lost by the beam in the foil is transferred to the electrons of the medium. Then, γ-rays are produced together with electrons of the energy below 50 eV that are called secondary electrons (SEs). The number of SEs ejected from the foil is proportional to the local beam intensity. The SEs are accelerated and focused by an electrostatic field towards an imaging device or a position-sensitive sensor, which provides the beam intensity and its position [28]. The secondary emission monitors are precise and reliable but relatively large and expensive. Moreover, frequent checks of these monitors are necessary because they may suffer from performance changes due to deposits or damages of the electrode surface.

### 3.7    Beam monitor calibration for scanned proton and carbon ion beams

The beam monitor (BM) for scanned charged particle beams measures the charge ($C$) collected during a given beam delivery and needs to be calibrated in units of particles per monitor unit (MU) (see Eq. (20)). The calibration has to be performed in a region with constant energy loss and usually the plateau region (i.e., the entrance region) is selected. The calibration is beam-energy dependent and must be validated (or determined) for all the available beam energies. The determination of the number of particles is based on a reference measurement of absorbed dose to water for a fixed set of energies. For example, a Farmer chamber having a calibration factor in terms of absorbed dose to water (reference beam quality is a Co[60] beam) is placed in a phantom in the centre of a homogeneous monoenergetic square field and irradiated with a regular grid of proton or carbon ion spots.

The calibration factor at energy $E$, $K(E)$, is defined as the number of particles $N$ per monitor unit MU, and is given by

$$K\left(E\right) = \frac{N}{\text{MU}} = \frac{D_{\text{meas}}}{\text{MU} \times S_{E(x)}} \Delta x \Delta y \,, \tag{20}$$

where $D_{\text{meas}}$ is the absorbed dose measured in the phantom, $S_{E(x)}$ is the mass stopping power of protons or carbon ions with the initial energy $E$ at the depth of measurement $x$, $\Delta x$ and $\Delta y$ are the spacings between two consecutive spots in the transversal direction (for example, 3 and 2 mm for protons and carbon ions, respectively). A pre-requisite for the application of Eq. (20) is that the scanned field delivers a homogeneous dose in the transverse plane. The measurement has to be performed at some representative energies $E_i$ (six at HIT [29] and nine at CNAO [30]) ranging from the minimum to the maximum, which corresponds to minimum and maximum depths of penetration in water. The delivered number of particles has to be set to a constant for each spot, corresponding to the selected number of MUs. The collected data, as shown in Fig. 26, can be fitted with a third-order polynomial curve, $K(E)$, that represents the BM calibration curve, one for each beam line and particle type and used for treatments by the software managing the dose delivery.

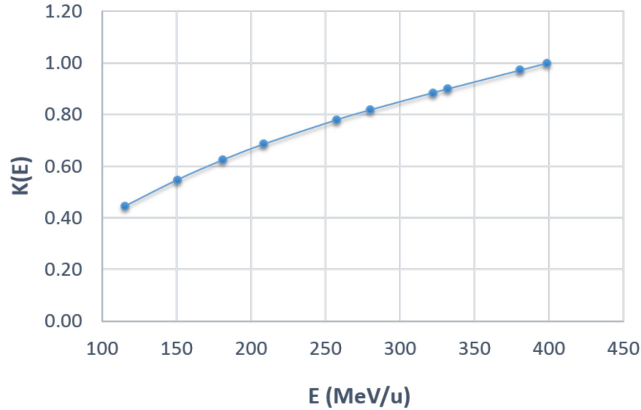

**Fig. 26:** Monitor calibration factor $K(E)$ as measured according to Eq. (20) for six different energies of the carbon ion beam.

## 3.8 Worldwide beam monitoring systems

### 3.8.1 The CNAO beam monitors

The CNAO dose delivery system [19] measures at a fixed frequency the number of delivered particles, the beam transversal position and dimension by means of five parallel-plate ionization chambers [31] filled with nitrogen. The monitor chambers are enclosed in two independent aluminium boxes: BOX1 and BOX2 (Fig. 27). The BOX1 contains an integral chamber (INT1) with a large-area anode for the measurement of the beam flux, followed by two chambers with the anodes segmented in 128 strips, 1.65 mm wide, respectively with vertical (StripX) and horizontal (StripY) orientations, which provide the measurement of the beam position and beam width projected along two orthogonal directions. The BOX2 contains a second integral chamber (INT2) followed by a chamber with the anode segmented in 32 × 32 pixels, each 6.6 mm wide (PIX). The measurements accomplished by the BOX2 detectors are beam flux, position and width.

The total water equivalent thickness of these chambers has been measured to be approximately 0.9 mm. The material budget interposed by the beam detectors is spread along approximately 20 cm and it is one of the main contributions to the beam lateral dispersion. To minimize the effect, the boxes are installed close to the patient, at approximately 70 cm from the isocentre.

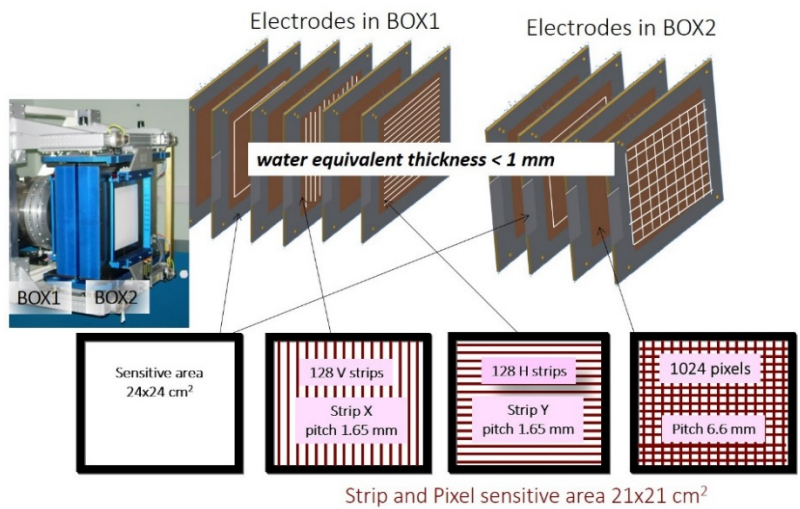

**Fig. 27:** Picture and sketch of the CNAO beam monitor detectors (BOX1 and BOX2)

The detector front-end readout is based on custom-designed boards, which host ASICs custom designed for this purpose. The chosen architecture and technology allow good sensitivity, with a minimum measured charge of 200 fC corresponding to a number of protons ranging from $7.2 \times 10^3$ at the energy of 62 MeV to $1.9 \times 10^4$ at the energy of 226 MeV. Similarly, the range of the number of carbon ions extends from $3.4 \times 10^2$ at the energy of 115 MeV/u to $7.7 \times 10^2$ at 399 MeV/u. The background current is limited to 200 fA; this ensures negligible error on the dose delivered to each spot.

The customized electronics is based on large-scale integration to provide many readout channels with uniform channel-to-channel behaviour, and allows a large segmentation of the active area.

The charge collected in the ionization chambers depends on the pressure and temperature of the gas, as well as on the high voltage across the gap. These quantities are monitored by a set of transducers installed in each box and controlled by a peripheral interface controller (PIC). Values are periodically read and checked by the PIC against pre-set values. The measured deviations are used to correct the gain of the chambers. Appropriate interlock procedures are activated whenever any of the values is outside the expected range.

The data, provided by the beam monitors in the CNAO local control room during the dose delivery, are shown in Fig. 28.

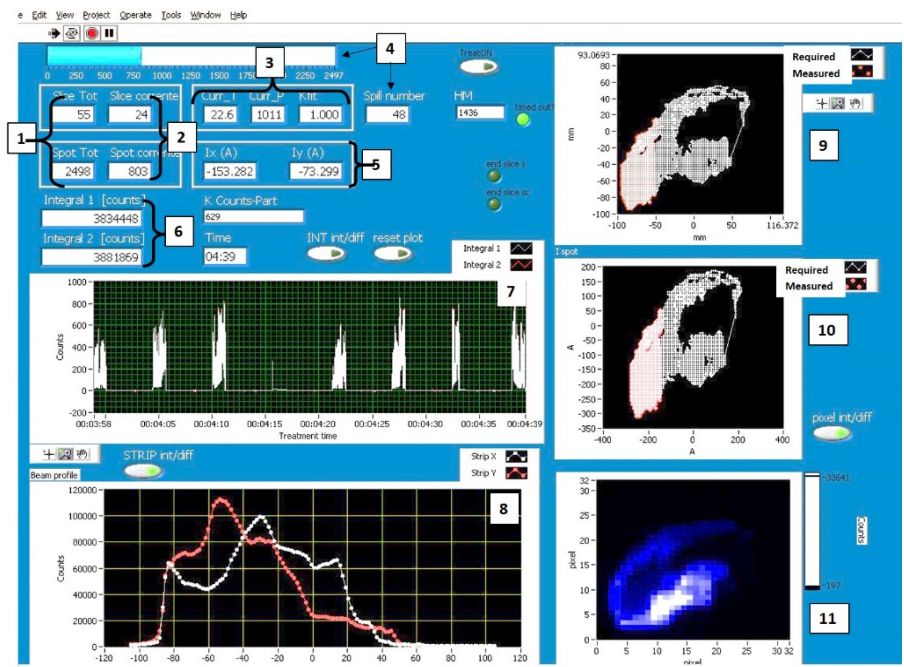

**Fig. 28:** Snapshot of the data published continuously on a monitor display in the CNAO local control room: (1) total number of slices and spots of this field; (2) number of the slice and the spot under treatment; (3) measured temperature, pressure and the flux correction factor; (4) number of the spill and delivery progress bar; (5) PS current set-points; (6) INT1 and INT2 total counts; (7) flux measured by INT1 and INT2; (8) StripX and StripY total counts; (9) spot positions in millimetres; (10) PS currents in amperes: measured (full dots covering partial field) and required (small dots covering the overall field); (11) 2D flux measured by PIX chamber.

### 3.8.2 The PSI Gantry 1 beam monitors [32]

The Paul Scherrer Institute Gantry 1 is equipped with four ionization beam chambers in the gantry nozzle to monitor the incident proton beam. The beam flux monitor 1 (Mon1) integrates the output ionization charge and determines the spot dose. When the integrated ionization charge from Mon1 reaches the expected value, the fast magnetic kicker switches off the beam. The main beam flux monitors, Mon1 and Mon2, are parallel-plate transmission ionization chambers (Fig. 29(a)) used to check the applied spot dose, while a strip ionization chamber (Fig. 29(b)) verifies the position and the width of the applied beam against the prescribed values.

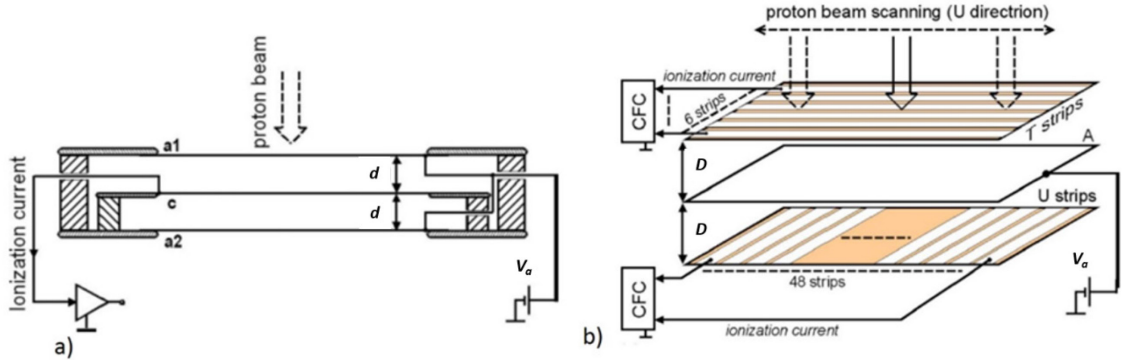

**Fig. 29**: (Reproduced from [32]) (a) Schematic representation of the PSI G1 parallel-plate ionization chamber (Mon1 and Mon2); $V_a$ (2000 V) is the applied voltage to the anodes (a1 and a2) and $d$ is the spacing between the anode and the cathode (c). Mon1 has $d = 0.5$ cm and Mon2 has $d = 1$ cm; (b) schematic representation of the PSI G1 strip ionization chamber.

The Mon1 and Mon2 chambers are filled with ambient air because these satisfy all the technical requirements and simplifies operation and maintenance compared to a closed gas system. The high-voltage planes (a1 and a2) are made of 20 μm thick Mylar foils single coated with aluminium with a thickness of less than 0.1 μm. The applied voltage ($V_a$) is 2000 V. The signal plane consists of a 20 μm thick aluminium foil. The chamber window covers the swept beam area (±9 cm) and is $23 \times 3$ cm$^2$ in size. Because the average duration of one spot is about 10 ms, the ionization charge collection time should be of the order of 0.1 ms in order to reach a precision of 1%. With the chamber gaps, Mon1 has an ion collection time $t_c$ of about 90 μs, and Mon2 has a $t_c$ of about 350 μs, having derived the collection time with the usual relationship:

$$t_c = \frac{d^2}{\mu . V_a},\tag{21}$$

where $V_a$ is the applied voltage, $d$ is the spacing between the high-voltage plane and the signal plane and μ is the mobility of charged particles in the electric field [in dry air, the average mobility μ is about 1.4 cm$^2$/(s V)]. The slower collection time of Mon2 is acceptable because it is used only as a back-up element and does not affect the precision of the dose delivery.

The strip chamber has a voltage plane made of 20 μm thick Mylar foil coated with aluminium layers (less than 0.1 μm thick). The signal planes consist of 20 μm thick Kapton foils coated with 4 mm wide aluminium strips. The spacing between two strips is 0.4 mm. The strip aluminium coating is about 1 μm thick. The 48 strips U and six strips T provide readings of the beam profiles in the horizontal and vertical directions. The counting gas is ambient air. By calculating the centre of gravity of the outputs of the strips hit by the beam, its position can be determined to approximately a tenth of a millimetre. The sums of the strip monitor outputs are also used as additional dose checks.

### 3.8.3    The Loma Linda University Medical Center beam monitors

At Loma Linda University Medical Center, a synchrotron produces accelerated protons to final energies between 70 and 250 MeV. The double scattering and the pencil beam scanning deliveries are available. A transmission ionization chamber (TIC) and a secondary electron emission monitor measure the integrated number of protons per spill or per treatment. Helium has been chosen as the filling gas because of helium's fast positive-ion mobility. As previously described, half of the intensity signal is due to the collection of electrons and half of the signal is due to the collection of positive ions. With field strength of 1 kV cm$^{-1}$ and an anode to cathode distance of 0.5 cm, the electron signal is detected quite fast, less than 10 ns, and the positive ions are collected linearly over time from 0 to 50 μs after the passage of the primary proton. The intensity signal is sampled at 50 kHz (every 20 μs) and processed (requiring another 10 μs) for determining if the beam position should be moved again [33].

The beam position and beam profile are monitored using three retractable multiwire ion chambers (MWICs). The wire chamber resolution is 2 mm. A $25 \times 25$ cm$^2$ ion chamber segmented into 400 square pads was placed after the range modulator to monitor the dose delivered to the target volume. The detector consists of a $20 \times 20$ array of pads (each $1.25 \times 1.25$ cm$^2$) from a thin sheet of gold-plated Kapton. Any of the pads in the central region of the pad plane can be used to monitor and control the dose delivered to the target and the remaining pads are also used to study the transverse dose distribution [26].

### 3.8.4    The Ion Beam Application (IBA) beam monitors for Pencil Beam Scanning (PBS) [34]

A $320 \times 320$ mm$^2$ parallel-plate ionization chamber composed of 15 Mylar foils separated by 5 mm air gaps has been developed to be installed in the IBA nozzles for PBS. The detector is composed of two identical units IC2 and IC3 with independent power supply and electronic acquisition set-up, for redundancy requirements. As sketched in Fig. 30, each unit is composed of five 2.5 μm Mylar electrodes coated on both sides with aluminium or gold. Three are connected to the high voltage while the two others are measurement electrodes at ground potential, one being used for dose measurement (uniform film) and the second one for beam position measurement (striped film) along one axis (horizontal for IC2 and vertical for IC3). Apart from the two units, three other films are connected to the ground to ensure the electrostatic pressure equilibrium. In addition, two thicker (25 μm) Mylar films are used to cover both entrance and exit windows. The whole chamber is 6.86 cm thick for a total water equivalent thickness of 187 μm.

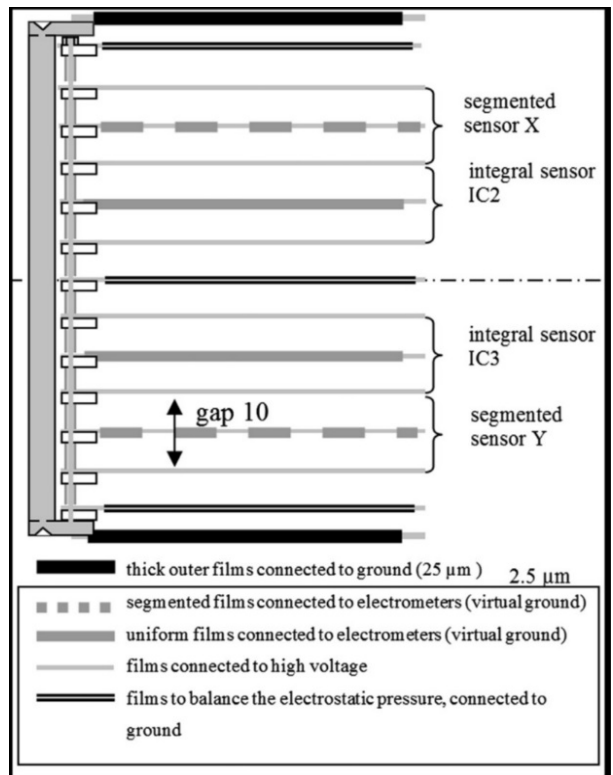

**Fig. 30:** (Reproduced from [34]) Vertical section of the IBA beam monitors for PBS

The position measurement layers are plain Mylar films covered by gold strips on both sides deposited by vacuum evaporation. There are 64 4.8 mm strips with a 5 mm step, oriented to give the $x$ and $y$ positions for IC2 and IC3, respectively. Dose measurement layers are single-layer aluminized Mylar foils covered in the plain side by a thin gold layer (200 nm). The dose is measured by integrating the signal collected on each film at a frequency of 2 kHz, allowing a time resolution of 500 μs.

### 3.8.5    The MD Anderson beam monitor system

The MD Anderson passive scattering nozzle [23], shown in Fig. 13, houses the following devices to monitor various aspects of the beam: the beam profile monitor, reference dose monitor, secondary dose monitor, primary dose monitor and multilayer Faraday cup.

The profile monitor measures the beam profile, the position of the beam centre and the beam width. The profile monitor sends a beam interlock signal that stops the beam if an out-of-tolerance condition is detected, i.e. if any measured value is unacceptably different from the pre-set value.

User-adjustable tolerance tables contain tolerances for a warning signal when the maximum tolerance value is approached and a pause signal when the maximum tolerance value has been reached. The primary and secondary dose monitors are segmented ionization chambers whose primary functions are to measure the monitor units (MUs) delivered and to terminate the treatment when the prescribed MUs have been delivered. For any specific treatment, the secondary dose monitor is set a few percent higher than the primary dose monitor so that the treatment is normally terminated by the primary dose monitor.

These two segmented ionization chambers also serve as beam flatness and symmetry monitors. They consist of a quadrant fan-shaped electrode and eight concentric circular electrodes. The quadrant fan-shaped electrode measures the beam centre, while the eight concentric electrodes measure the lateral beam distribution.

## 4    The dose delivery control concept

The dose delivery instrumentation has a crucial role within the complex architecture of a facility for charged particle radiotherapy. Their advanced control software and IT systems, mainly in the context of high-speed scanning techniques, can have a significant level of risk as a side effect.

Therefore, DD instrumentation needs to be integrated in control (or interlock) systems, which immediately stop the irradiation as a reaction to any condition leading to a potential hazard.

At CNAO, for example, the safety of the treatment mainly relies on two interlock systems: the patient interlock system (PIS) and the safety interlock system (SIS). These systems collect any error conditions and either force the immediate interruption of the beam delivery or inhibit the operations as long as the conditions persist. The PIS is dedicated to the patient safety by acting on the beam chopper to interrupt the treatment when an interlock occurs. It manages both short interruptions (a few seconds) and treatment termination and recovery. Additionally, for recovery purposes, one battery back-up device, called a dose delivered recovery system, continuously receives, stores and displays the last treated slice and the spot of each slice during irradiation [19]. For passive scattering systems, the backed-up storage of measured monitor units is also recommended.

At PSI, a rigorous separation has been implemented between the tasks and responsibilities of the machine as a beam delivery system and a user who asks for a beam with specified characteristics and decides whether the beam is accepted for treatment. A machine control system (MCS) controls the accelerator's and beam line's performance. Then, each treatment area has its own therapy control system (TCS) [35].

## Acknowledgements

We wish to thank F. Marchetto and R. Sacchi for useful comments and a careful review of this document.